\documentclass[11pt]{article} 
\pdfoutput=1
\usepackage{amssymb,graphicx} 
\usepackage{epstopdf}
\usepackage{amsmath,amsfonts}
\usepackage{epsfig} 
\usepackage{graphicx,graphics}
\usepackage[dvipsnames]{xcolor}
\usepackage[utf8]{inputenc}
\usepackage{hyperref}

\begin{document} 

\sloppy

\title{\bf Gravitational shine of dark domain walls}
\author{E.~Babichev$^{a,b}$, D.~Gorbunov$^{c,d}$, S.~Ramazanov$^e$, A.~Vikman$^{e}$\\
\small{\em $^a$Universit\'e Paris-Saclay, CNRS/IN2P3, IJCLab, 91405 Orsay, France}\\
\small{\em $^b$Institute for Theoretical and Mathematical Physics,}\\
\small{\em Lomonosov Moscow State University, 119991 Moscow, Russia}\\
\small{\em $^c$Institute for Nuclear Research of the Russian Academy of Sciences, Moscow 117312, Russia}\\ 
\small{\em $^d$Moscow Institute of Physics and Technology, Dolgoprudny 141700, Russia}\\
\small{\em $^e$CEICO, FZU-Institute of Physics of the Czech Academy of Sciences,}\\
\small{\em Na Slovance 1999/2, 182 21 Prague 8, Czech Republic}
}

{\let\newpage\relax\maketitle}

{\let\newpage\relax\maketitle}

\begin{abstract}

Cosmic domain walls are harmless, provided that their tension decreases with expansion of the Universe. This setup can be realized, if the scale of spontaneous symmetry breaking 
is induced dynamically through the
interaction with hot primordial plasma. 
In that case, the domain wall tension can attain large values in the early Universe without any conflict with observations. 
Owing to the large initial tension, these topological defects may serve as a powerful source of gravitational waves. We make a preliminary estimate of the gravitational wave spectrum and argue that it is distinct from the spectrum produced by other sources, in particular 
by domain walls of a constant tension. The resulting gravitational wave signal is in the
range accessible by Einstein Telescope, DECIGO, TianQin, LISA, IPTA, or SKA, if the field constituting the domain walls is very feebly coupled
with hot primordial plasma and has tiny self-interactions. In
particular, one can consider this field for the role of Dark
Matter. We discuss various Dark Matter production mechanisms and properties of the emitted gravitational waves associated with them. We
find that the conventional freeze-out and freeze-in mechanisms 
lead to large and perhaps unobservable frequency of gravitational
waves. However, the Dark Matter production is also possible at the second
order phase transition leading to the domain wall formation or at the
inverse phase transition, when the domain walls get dissolved
eventually.  
In both cases, there is essentially no lower bound on the frequency of
emitted gravitational waves.

\end{abstract}

\section{Introduction} 

Spontaneous breaking of discrete symmetries in the early Universe
generally leads to formation of domain walls~\cite{Zeldovich:1974uw,
  Kibble:1976sj}. The latter are often regarded as unwelcome guests in
cosmology: their energy density redshifts too slowly relative to the
energy density of the surrounding matter.  As a result, domain walls
overclose the Universe~\cite{Zeldovich:1974uw}, unless their tension
(wall mass per unit area) is constrained as $\sigma_{wall} \lesssim
\left(\mbox{MeV} \right)^3$. Known ways of curing the problem involve
an explicit breaking of discrete symmetries~\cite{Zeldovich:1974uw,
  Gelmini:1988sf}, biased initial conditions~\cite{Coulson:1995nv},
embedding discrete symmetry into a continuous
group~\cite{Lazarides:1982tw}, symmetry non-restoration at high
temperatures~\cite{Dvali:1995cc} or instead symmetry restoration at
substantially low temperatures~\cite{Vilenkin:1981zs,
  Vilenkin:2000jqa}. Among all, the latter scenario is of our major
interest in this work. It assumes that vacuum expectation values $\pm
\eta$ of the field constituting the domain walls are not constant in
the expanding Universe, but
being induced dynamically through the interaction with the
primordial plasma of temperature $T$, i.e., $\eta \propto T$. Consequently, the wall
tension $\sigma_{wall} \propto \eta^3$ evolves as
\begin{equation}
\label{main}
\sigma_{wall} (t) \propto T^3 (t) \; .
\end{equation}
Therefore, the energy density of the domain wall network drops faster
than the energy density of radiation, and thus the standard
cosmological evolution is not affected. Eventually one expects the
inverse phase transition to take place~\cite{Weinberg:1974hy} and the
domain walls vanish completely. In the present work we investigate
phenomenological manifestations of domain walls with
tension~\eqref{main} elaborating on and extending the results obtained
in our earlier work~\cite{Ramazanov:2021eya}.

Such domain walls can be realized in a simple renormalizable setup
involving two scalar fields: a real singlet scalar $\chi$ 
possessing the discrete $Z_2$-symmetry and a scalar multiplet $\phi$, which
is assumed to be in thermal equilibrium with the surrounding
plasma. The two fields 
interact through the scalar portal coupling $\sim g^2 \chi^2 \phi^{\dagger} \phi$. At high
enough temperatures and for the proper choice of the coupling constant
$g^2$, the interaction between the two fields spontaneously breaks
$Z_2$-symmetry, and domain walls form.  
There are two possible
scenarios of the late time domain wall evolution. They are either dissolved completely, if the inverse phase transition occurs at 
later time, or the tension may stabilize at a constant value if no inverse phase transition happens. This value should be below the existing observational upper bound $\sim$ (MeV)$^3$.

Domain walls with tension~\eqref{main} have a negligible effect on the
cosmological evolution of the Universe. Nevertheless, their
phenomenological properties can be rather prominent. Owing to the high
initial tension, the domain walls emit potentially observable
gravitational waves (GWs)~\cite{Hiramatsu:2013qaa}\footnote{See
Refs.~\cite{Gleiser:1998na, Sakharov:2021dim}, which consider GWs
produced in the non-spherical collapse of closed domain walls.}. The
origin of GWs is as follows.
Domain walls move inside the cosmological horizon, cross each other and merge, until the net of domain walls 
enters the scaling regime with one or a few domain walls per horizon
volume~\cite{Press:1989yh, Garagounis:2002kt, Leite:2012vn}. 
During the evolution the domain wall network develops a large time-dependent
quadrupole moment, which sources GW emission. Because of the
time-decreasing tension, most GWs are emitted at the earliest possible
times, when domain walls just enter the scaling regime. This behavior is in a
sharp contrast with the scenario that involves constant-tension domain walls, 
when the peak of GW emission corresponds to late times, right
before the domain wall dissolution. Therefore, one may expect a rather
distinct spectrum of GWs produced in our scenario.  
We discuss this point in Section~4.

If GWs are emitted by the domain walls formed at the radiation-dominated
stage, their characteristics are fully determined by the model
parameters: the portal coupling constant $g^2$ and the self-interaction
constant of the field $\chi$.  In particular, for the domain wall formation temperatures $T_i \lesssim
5\cdot 10^{9}~\mbox{GeV}$, the peak frequency and the amplitude of GWs falls into the range
accessible at the future detectors. If domain walls are formed during
preheating, the GW characteristics are generically determined not only
by the model parameters, but also by the reheating temperature, maximal
temperature of the Universe, equation of state of the dominant matter,
and/or parameters describing the temperature evolution. We show,
however, that for fixed model parameters, in general, the signal is stronger if GWs are emitted at the radiation-domination. Nevertheless, if a stage with sufficiently stiff equation of state (e.g., kination) took place after inflation, one can expect a stronger signal of GWs produced at preheating.

In our setup, observable GWs correspond to very weak couplings of the
underlying model. Therefore, the field $\chi$ constituting domain
walls can be considered for the role of Dark Matter (DM). Naturally, DM is produced at the phase transition, leading to the domain walls formation\footnote{See also the recent
works~\cite{Baldes:2018emh, Azatov:2021ifm, Nakayama:2021avl} and
references there discussing other ways of DM production at the first
or second order phase transition and the associated GW signal.}  
when the field $\chi$ moves from zero value to the minima of the broken phase. 
The field $\chi$ reaches the new minimum with a large velocity, and starts to oscillate. These
oscillations survive till present and feed into DM, unless the field
$\chi$ has decay channels. Remarkably, in the range of parameters,
which leads to observable GWs, the temperature at the onset of the DM
phase is very low, in the ballpark of $1~\mbox{keV}$.  
This suggests an interesting interplay between observations of GWs and large scale structure
observations. It should be stressed that an efficient DM production is possible, even if
oscillations of the field $\chi$ quickly relax to zero in the
spontaneously broken phase. In that case, one can achieve the right DM
abundance at the inverse phase transition. This mechanism, which is
also consistent with potentially observable GWs, has been already
discussed in our earlier work~\cite{Ramazanov:2021eya} and we review
it in Subsection 6.2.  Finally, we discuss the DM production by the freeze-in mechanism, which operates for a fixed value of
the portal constant translating into a certain value of the peak GW
frequency. However, the latter turns out to be very high (it is even higher in the case of the freeze-out mechanism) challenging
the search for the associated GW signal.

\section{$Z_2$-symmetry breaking at high temperatures}

Let us consider the following Lagrangian describing a real scalar singlet $\chi$:
\begin{equation}
\label{basic}
{\cal L}_{\chi}= \frac{(\partial {\chi})^2}{2} +\frac{\xi R \chi^2}{2}+ \frac{g^2 \chi^2 \phi^{\dagger} \phi}{2} -\frac{\lambda \chi^4 }{4} \; ,
\end{equation}
where $\lambda$ is the self-interaction coupling constant and $g^2$ is
the portal coupling constant to a scalar multiplet $\phi$ being in
equilibrium with the primordial plasma. Apart from that, we do not
make particular assumptions about the field $\phi$, unless otherwise
stated. The coupling to the Ricci scalar $R$ controlled by a
dimensionless constant $\xi$ should be generically anticipated on the
base of renormalization considerations in the curved space-time~\cite{Birrell}. At the
moment, we do not include the bare mass term for the field $\chi$ in
the action~\eqref{basic}. The part of phenomena discussed in this work
occur at very large temperatures, when bare masses are negligible. In
that case, the Lagrangian~\eqref{basic} enjoys a nice property of
scale-invariance, which is slightly broken by the inevitable coupling
to gravity.  Bare mass terms come into play at low temperatures and
they are crucial for the DM phenomenology of the model, which we
consider later in this work.

For $\xi \gtrsim 1$\footnote{Note that the parameter $\xi$ is practically unconstrained from above. Indeed, the strong coupling scale $\Lambda \sim \frac{M_{Pl}}{\xi}$ induced by the non-minimal coupling of the field $\chi$ to gravity is well above all the energy scales considered in the present work, at least for not exponentially large $\xi$.}, the field $\chi$ has a super-Hubble mass during
inflation and hence quickly relaxes to zero: 
\begin{equation}
\label{backzero}
\chi =0  \; .
\end{equation}
This should be set as the initial condition for the evolution of the field $\chi$ at the hot Big Bang stage. The condition~\eqref{backzero} and thus the inequality $\xi \gtrsim 1$ are necessary for domain wall formation (see also the discussion after Eq.~\eqref{Mthermal2}). As the Universe reheats and temperatures reach large values, the interaction with 
the field $\phi$ comes into play, while the coupling to the Ricci scalar can be neglected. Following our earlier work~\cite{Ramazanov:2021eya}, we assume that the portal coupling constant $g^2$ is positive: 
\begin{equation}
\label{positive}
g^2 >0 \; .
\end{equation} 
Consequently, thermal fluctuations of the field $\phi$ feed into the non-zero expectation value of $\chi$:
\begin{equation}
\label{vacthermal}
\eta^2 (T) =\frac{g^2 \langle \phi^{\dagger} \phi \rangle_T}{\lambda} \; .
\end{equation}
Thus, the effective potential $V_{eff} (\chi)$ accounting for the one-loop thermal corrections, exhibits spontaneous symmetry breaking\footnote{See also recent studies~\cite{Chai:2020onq, Chai:2020zgq, Chai:2021djc} of conformal field theories exhibiting spontaneous symmetry breaking at all finite temperatures.}:
\begin{equation}
\label{spontaneous}
V_{eff}=\frac{\lambda \cdot \left(\chi^2-\eta^2 (T) \right)^2 }{4} \; .
\end{equation}
Compared to Eq.~\eqref{basic}, here we added a $\chi$-independent term, which does not affect equations of motion of the field $\chi$.
At high temperatures, the particles $\phi$ are relativistic and one has
\begin{equation}
\label{thermalaverage}
\langle \phi^{\dagger} \phi \rangle_{T} \approx \frac{NT^2}{12} \; ,
\end{equation}
where $N$ counts the number of degrees of freedom entering the scalar multiplet $\phi$. Using Eqs.~\eqref{vacthermal} and~\eqref{thermalaverage}, we obtain
\begin{equation}
\label{tension}
\eta^2 (T) \approx \frac{Ng^2 T^2}{12 \lambda}  \; .
\end{equation}
Our choice of sign for $g^2$ imposes an important limitation on the
model parameter space:
\begin{equation}
\label{runaway}
\beta \equiv \frac{\lambda}{g^4} \geq \frac{1}{\lambda_{\phi}}  \; ,
\end{equation}
where $\lambda_{\phi}$ is the self-interaction coupling constant of
the field $\phi$. The condition~\eqref{runaway} guarantees absence of
runaway solutions in the field space $(\chi, \phi)$ and hence protects
vacuum stability. Another limit on $\beta$ follows from the loop
correction to the self-coupling constant $\lambda$ triggered by the
portal interaction, i.e., $\delta \lambda \sim N g^4/(4\pi^2)
$. Baring fine-tuning, $\lambda$ cannot be smaller than the loop
correction $\delta \lambda$ and hence
\begin{equation}
\label{loop}
\beta \gtrsim \frac{N}{4\pi^2} \; .
\end{equation}
In practice, we will always assume
\begin{equation}
\beta \geq 1 \; ,
\end{equation}
which is consistent with $\lambda_{\phi} \leq 1$ according to Eq.~\eqref{runaway}. Values of $\beta$ close to the lower bound are most interesting from the viewpoint of GW phenomenology, which we discuss in Section~4.

In the end of this Section, let us review how the expression~\eqref{spontaneous} follows from the one-loop temperature approach~\cite{Dolan:1973qd}. Keeping only diagrams, which involve the field $\phi$ in the propagator, one obtains
\begin{equation}
V_{1-loop} (\chi) \approx \frac{NT^4}{2\pi^2} \int^{+\infty}_0 dx ~x^2~ \ln \left[1-\mbox{exp} \left(-\sqrt{x^2+\frac{m^2_{\phi}}{T^2}} \right) \right] \; . 
\end{equation}
The leading order non-trivial term in the limit of large temperature $T \gg m_{\phi}$ is given by 
\begin{equation}
V_{1-loop} (\chi) \approx \frac{NT^2 m^2_{\phi}}{12} \; .
\end{equation}
From $m^2_{\phi}=-g^2\chi^2+....$, where the ellipsis corresponds to
the terms independent of the field $\chi$, we obtain
Eq.~\eqref{spontaneous}. Here let us also note that the full mass
squared of the field $\phi$ is always positive, despite that it
receives the negative correction from the interaction with the field
$\chi$. Indeed, the thermal part of the field $\phi$ mass squared
fullfils $\delta m^2_{\phi} (T) \gtrsim N\lambda_{\phi} T^2/12$, which is
always larger than the negative contribution due to the field $\chi$,
once the condition~\eqref{runaway} is obeyed. Therefore, unlike
$\chi$, the field $\phi$ is in the unbroken phase at high
temperatures.

\section{Formation of domain walls at radiation dominated stage}

As the field $\chi$ rolls from the origin to the minimum of the effective potential~\eqref{spontaneous}, domain walls rise. 
This formation process is conceptually different 
in our scenario as compared to the standard case. Indeed, in the standard scenario, the relevant field is initially set to zero due to a large thermal mass. As the Universe temperature decreases, 
 so does the thermal mass. Eventually the latter becomes smaller than the absolute value of the bare tachyonic mass, 
 and the field starts to roll, forming the domain walls. In our case, on the contrary it is the thermal mass\footnote{Recall that the field $\chi$ is not in thermal equilibrium with the primordial plasma.} of the field $\chi$, which is tachyonic: 
 \begin{equation}
   \label{Mthermal2}
M^2_{thermal}=-\lambda \eta^2 (T) \; .
\end{equation}
This thermal mass is responsible for spontaneous breaking of $Z_2$-symmetry. The roll of the field $\chi$ starts, when $|M_{thermal}|$ overcomes the Hubble friction. The field $\chi$ has an equal probability to pick positive and
negative values in different Hubble patches. The neighbouring patches
containing different vacuum values of the field $\chi$ become
separated by the domain walls. Here it is crucial that initially the background field $\chi$ is zero, see Eq.~\eqref{backzero}, which is the case if $\xi \gtrsim 1$ in Eq.~\eqref{basic}. Otherwise, for $\xi \ll 1$ the field $\chi$ develops 
large superhorizon perturbations during inflation shifting its background value to $\chi \neq 0$. This creates a strong bias towards vacuum values with a particular sign (correlated with the sign of initial background $\chi$), and 
no domain walls are formed.

According to the above discussion, the time $t=t_h$, when the roll
towards the symmetry breaking phase starts, is defined from the
equality 
\begin{equation}
\label{beginning}
|M_{thermal}(T_h) | = \frac{N^{1/2}gT_h}{\sqrt{12}} \simeq H_h \; .
\end{equation}
Here $T_h$ and $H_h$ are the temperature and the Hubble rate at the time $t_h$, respectively. In this Section, we assume that domain walls are formed at the radiation-dominated stage, so that the Hubble rate is given by 
\begin{equation}
\label{hubble}
H =\sqrt{\frac{\pi^2 g_{*} (T)}{90}} \cdot  \frac{T^2}{M_{Pl}} \; ,
\end{equation}
where $M_{Pl} \approx 2.44 \times 
10^{18}$\,GeV is the reduced Planck mass, and 
$g_* (T)$ is the number of relativistic degrees of freedom at the temperature $T$. In what follows, we ignore the difference between $g_* (T)$ and the number of degrees of freedom entering the entropy density. The rolling phase has a finite duration; therefore, 
the temperature at the time $t_i$, when the roll finishes, is below $T_h$ defined by Eq.~\eqref{beginning}, i.e., 
\begin{equation}
\label{initial}
T_i = \frac{T_h}{\sqrt{B}} \simeq  \frac{N^{1/2}gM_{Pl}}{ \sqrt{B g_* (T_i)}} \; .
\end{equation}
Here  
\begin{equation}
  \label{B-coeff}
B \equiv \frac{T^2_h}{T^2_i} \gtrsim 1 \; ,
\end{equation}
is the \emph{rolling} parameter, which takes into account the time required for the
field $\chi$ to reach the broken minimum. We derive the coefficient
$B$ below in this Section. Note that the domain walls start growing before the time $t_i$, when the second order phase transition is accomplished. However, their tension is yet too small to lead to any phenomenological consequences. Thus, from the viewpoint of any practical applications, 
we can consider $t_i$ (when the Universe temperature is $T_i$) to be
the moment of domain wall formation. The dynamics of the domain wall
growth between $t_h$ and $t_i$ is important, however, for finding the
exact spectrum of the GWs in the model. This task in any case requires
numerical simulations which is beyond the scope of this paper.

The key feature of our scenario is a non-constant domain wall tension,
\begin{equation}
\label{surface}
\sigma_{wall} = \frac{2\sqrt{2\lambda} \eta^3 (T)}{3} \; ,
\end{equation}
which weakens with time as the power of the temperature, $\sigma_{wall} (T) \propto T^3$, see Eq.~\eqref{tension}. 
As a result, the domain walls are melting in the surrounding plasma, and hence do not have any appreciable effect on the cosmological evolution~\cite{Vilenkin:1981zs}. Let us take a closer look at this important point. The domain wall network settles to a scaling regime with 
one (a few) wall(s) in a Hubble patch~\cite{Press:1989yh,
  Garagounis:2002kt, Leite:2012vn}. Thus we estimate the domain wall
mass inside the Hubble volume as
\begin{equation}
\label{wallmass}
M_{wall} \sim \frac{\sigma_{wall}}{H^2} \; .
\end{equation}
Consequently, the estimate of the domain wall energy density is given by 
\begin{equation}
\rho_{wall} \sim M_{wall} H^3 \sim \sigma_{wall} H \; .
\end{equation}
Substituting Eqs.~\eqref{tension},~\eqref{hubble}, and~\eqref{surface}, we obtain 
\begin{equation}
\rho_{wall} \sim \frac{g^{1/2}_{*} (T) ~N^{3/2}  g^3 T^5}{100\lambda M_{Pl}}\; .
\end{equation}
We express the portal constant $g$ through the parameter $\beta$ defined by Eq.~\eqref{runaway} and the Planck mass through the temperature $T_i$ given by Eq.~\eqref{initial}. Assuming $g_* (T) \approx g_* (T_i)$, we get
\begin{equation}
\rho_{wall} \sim \frac{ N^2 T^5}{100 \sqrt{B} \beta T_i} \; .
\end{equation}
Hence, the contribution of domain walls to the total budget of the
Universe during radiation dominated stage is estimated as 
\begin{equation}
  \label{omega-wall}
\frac{\rho_{wall}}{\rho_{rad}} \sim \frac{N^2}{30 \sqrt{B} g_{*} (T) \beta} \cdot \frac{T}{T_i} \; ,
\end{equation}
where we substituted $\rho_{rad}=\pi^2 g_*(T)T^4/30$. The above ratio is always small provided that the condition~\eqref{runaway} is obeyed and $N$ is not very large. This is in contrast to the scenario with constant tension domain walls, which
overclose the Universe, unless their tension is extremely low or there
is an explicit breaking of $Z_2$-symmetry. 

As it follows from Eq.~\eqref{omega-wall}, the fractional energy density kept in the domain
walls depends on the coefficient
$B$ defined by Eq.~\eqref{B-coeff}. 
To derive it, we make the following simple observation.
The redefined field $\tilde{\chi} =\chi a$, where $a(t)$ is the cosmological scale factor, effectively evolves in the flat space-time: 
\begin{equation}
\label{flat}
\tilde{\chi}^{''}-\partial^2_i \tilde{\chi} + \tilde{M}^2_{eff}  \tilde{\chi}=0 \; ,
\end{equation}
where the prime $'$ denotes the derivative with respect to the
conformal time $\tau$ defined via $dt=ad\tau$, and the rescaled
squared effective mass is 
\begin{equation}
\tilde{M}^2_{eff} \equiv  M^2_{eff} \cdot a^2 (\tau)   =3\lambda
\tilde{\chi}^2+\tilde{M}^2_{thermal} \;. 
\end{equation}
At small initial values of the field $\chi$, the term $\sim \lambda \tilde{\chi}^2$ in the
above squared mass is irrelevant, and the field $\tilde{\chi}$
experiences a tachyonic instability with a rate described by
$|\tilde{M}_{eff}| \approx |\tilde{M}_{thermal}|$, which is constant
in time as long as $g_*$ is constant, see Eqs.~\eqref{tension} and~\eqref{Mthermal2}:
\begin{equation}
  \label{constant-mass}
\tilde{M}^2_{thermal}\equiv a^2 M^2_{thermal}=\frac{Ng^2 T^2a^2}{12} =\text{constant} \; .
\end{equation}
Extrapolating it all the way down to the moment, when $\chi$ reaches the minimum, we can write the solution for large wavelength modes of
the field $\tilde{\chi}$ assuming the initial condition $\tilde{\chi}'
\simeq 0$ at the conformal time $\tau_h$:
\begin{equation}
\label{greatsolution}
\tilde{\chi} (\tau) \simeq \frac{1}{2} \tilde{\chi}_h \left[ \mbox{exp} \left( |\tilde{M}_{thermal} | \cdot (\tau-\tau_h) \right)+ \mbox{exp} \left(-|\tilde{M}_{thermal} | \cdot (\tau-\tau_h) \right) \right] \; .
\end{equation}
Here $\tilde{\chi}_h\equiv \chi_h a_h$, and $\chi_h$ is the value of
the field $\chi$ at the onset of the roll at $\tau_h$. Its possible
origin will be discussed shortly. At the end of the roll, the field is at the minimum of the broken potential $\chi_i=\eta_i$, and therefore
$\tilde{\chi}_{i} =\eta_i \cdot a_i$, which equals $\eta_h a_h$, since
$\eta\propto T$ and $T\propto 1/a$. Now, restoring the original variables, and assuming $\tau_i\gg \tau_h$, we arrive at
\begin{equation}
| M_{thermal}(t_h) | \cdot a_h \tau_i \simeq \ln \left(\frac{2\eta_h}{\chi_h} \right)   \; .
\end{equation}
Finally, using Eq.~\eqref{beginning} and expressing the Hubble parameter
via conformal time and scale factor, $|M_{thermal, h} | \simeq H_h
=1/(a_h \tau_h)$, we obtain (recall that  $a\propto\tau\propto 1/T$ at
the radiation domination):
\begin{equation}
\label{A}
B \equiv \left(\frac{T_h}{T_i} \right)^2 \simeq \left( \frac{\tau_i}{\tau_h} \right)^2 \simeq \ln^2 \left(\frac{2\eta_h}{\chi_h} \right)\; .
\end{equation}
It is crucial that $\chi_h \neq 0$ in order to trigger the phase transition. 
Naturally, $\chi_h$ is associated with the superhorizon perturbation of the field $\chi$. Naively, one expects the perturbation $\chi_h$ to be extremely small, because the field $\chi$ is assumed to have a super-Hubble mass during inflation due to the coupling to the Ricci scalar. 
However, the same non-minimal coupling to gravity generically amplifies perturbations of the field $\chi$ during preheating. The reason is that the Ricci scalar undergoes  oscillations, which reflect oscillations of an inflaton. 
As a result, the field $\chi$ experiences the tachyonic instability during the periods of time, when the Ricci scalar 
is positive~\cite{Markkanen:2015xuw, Fairbairn:2018bsw,
  Bassett:1997az, Felder:2000hj}. This leads to the amplification of
the fluctuation $\chi_h$, which may easily reach the values as
large as $\chi_h \simeq \eta_h$, at least for some range of
$\xi$. In that case, we have $B \simeq 1$. Generically, however, $B$ can be much larger; in what follows we keep it as a free parameter varying in the range $B \simeq 1-100$.

Let us note in passing one interesting feature of the above analysis. 
As it follows from Eq.~\eqref{flat}, neglecting gravitational
backreaction the evolution of domain walls in our model is equivalent
to the evolution of domain walls with a constant tension in the Minkowski spacetime. 
This property can be traced back to the scale-invariance of the underlying model. We believe, this may greatly simplify numerical analysis of domain wall evolution. 
See Ref.~\cite{Emond:2021vts} for the analogous observation in the case of melting cosmic strings.

We conclude with two important remarks. From the viewpoint of GW
emission, one is more interested in the time, when domain walls enter
the scaling regime rather than their formation time. The reason is that GW emission is non-efficient before the system settles to the scaling regime~\cite{Hiramatsu:2013qaa}. These times do not necessarily coincide. As it follows from the analysis of Ref.~\cite{Hiramatsu:2013qaa}, 
the scaling regime starts, when the domain wall width given by
 \begin{equation}
 \label{deltaw}
 \delta_{w} \simeq \frac{\sqrt{2}}{\sqrt{\lambda}\eta } \; ,
 \end{equation}
becomes small relative to the horizon size $\sim 1/H$, i.e., 
\begin{equation}
\label{scaling}
\delta_{w} \simeq 0.1 H^{-1} \; .    
\end{equation}
In our case, we find, using Eqs.~(\ref{hubble}),~(\ref{surface}), and (\ref{deltaw}),
\begin{equation}
\delta_{w, i} H_i \simeq \frac{1.5}{\sqrt{B}} \; .
\end{equation}
Therefore, for $B \simeq100$, the condition~\eqref{scaling} is fulfilled, once 
the field $\chi$ has reached its minimum. Consequently, for such large $B$ one expects that the scaling regime starts pretty much immediately upon the time of domain wall formation $t_i$. On the other hand, for $B \lesssim 100$ there can be a substantial 
time gap between $t_i$ and the moment, when the scaling regime is established. This time gap should be taken into account, when studying GW emission. 
Our arguments here are based on the analysis performed for domain walls with a constant tension~\cite{Hiramatsu:2013qaa}.
As it is mentioned above, in our case the domain walls effectively
evolve in the flat spacetime, and the relation~\eqref{scaling}
defining the onset of the scaling regime can be altered. Clarifying
this issue requires proper numerical simulations.

Finally, in the present Section we implicitly assumed that domain walls move with relativistic velocities $v\sim 1$; otherwise, settling of the system to the scaling regime would be further delayed. We argue below that this is indeed the case in our scenario. Generically, the motion of domain walls can be slowed down due to the 
interaction with the thermal particles $\phi$. However, in a very weakly coupled regime we are interested in, the resulting thermal friction force per unit wall area $F_{d} \sim M^2_{eff} (T) \cdot T^2 \cdot v$
is negligible compared to the force per unit wall area $F_{wall} \sim \sigma_{wall}/(vt)$ caused by the domain wall tension. Here $M_{eff} (T) =\sqrt{2\lambda} \eta (T)$ is the effective mass of $\chi$ in the broken phase. We took the estimates for $F_{d}$ and $F_{wall}$ from Ref.~\cite{Vilenkin:2000jqa}, see the Chapter~13 there. Indeed, the condition $F_{d} \lesssim F_{wall}$ is equivalent to the constraint on the velocity 
\begin{equation}
v \lesssim \left(\frac{g^{1/2}_* (T)\cdot \sigma_{wall} (T)}{3 M^2_{eff} (T) \cdot M_{Pl}} \right)^{1/2} \; .
\end{equation}
Using Eqs.~\eqref{tension},~\eqref{hubble},~\eqref{initial}, and~\eqref{surface}, we can rewrite this constraint as 
\begin{equation}
v \lesssim \frac{1}{5} \cdot \sqrt{\frac{N}{B}} \cdot \frac{1}{\beta^{1/2} g} \cdot \left(\frac{a_i}{a(t)} \right)^{1/2} \; .
\end{equation}
We will see in the next Section that observable GWs correspond to $g \lesssim 10^{-8}$ and $\beta \lesssim 100$; furthermore, they are mostly emitted at the times close to $t_i$. For these values, there is clearly no bound on the domain 
wall velocity, which can be as large as $v \sim 1$.

\section{Gravitational waves from domain walls}

Now let us turn to the discussion of GWs produced by the network of melting walls. The power of gravitational radiation can be roughly estimated using the Einstein quadrupole formula:
\begin{equation}
P \sim \frac{\dddot{Q}_{ij} \dddot{Q}_{ij}}{40\pi M^2_{Pl}}  \; .
\end{equation}
In the scaling regime, when there is one or a few domain walls in a Hubble patch of the size $H^{-1}$, the quadrupole moment $Q_{ij}$ is related to the mass of the wall $M_{wall}$ inside this patch by 
\begin{equation}
|Q_{ij}| \sim \frac{M_{wall}}{H^2} \; .
\end{equation} 
Using Eqs.~\eqref{surface} and~\eqref{wallmass}, we estimate the energy density of GWs produced at the time $t$ as
\begin{equation}
\label{rough}
\rho_{gw} \sim (P \cdot t) \cdot H^3 \sim \frac{\sigma^2_{wall}}{40\pi M^2_{Pl}} \sim \frac{\lambda \eta^6}{40\pi M^2_{Pl}}\; .
\end{equation}
Applying the above formula to our case and substituting Eq.~\eqref{tension}, we get
\begin{equation}
\rho_{gw} \sim \frac{5 N^3g^6 T^6 }{10^6 \lambda^2 M^2_{Pl}} \; .
\end{equation}
One observes that most energetic GWs are emitted at high temperatures
$T \simeq T_i$ close to the time of the domain wall formation. This is a direct consequence of the temperature-dependent tension $\sigma_{wall} (T)$. 
Were the tension $\sigma_{wall}$ constant, we would end up with the constant energy density of GWs.

We rely on the results of lattice simulations of Ref.~\cite{Hiramatsu:2013qaa}, which investigates the properties of 
GWs in the scenario with constant tension domain walls. 
One of the main conclusions following from Ref.~\cite{Hiramatsu:2013qaa} is that the rough estimate~\eqref{rough} matches well results of numerical simulations. Note that the estimate~\eqref{rough} is equally applicable to domain walls 
with a constant and time-dependent tension, if the latter changes at
the rate comparable or lower than the Hubble rate (which is the case
at hand, once the field $\chi$ settles at the minimum). 
Therefore, we expect the results of Ref.~\cite{Hiramatsu:2013qaa} to hold in our scenario at the qualitative level. The important qualification concerns determination of the GW spectrum, which will be discussed 
below in this Section.

We quantify the energy density of GWs in terms of its fraction:
\begin{equation}
\Omega_{gw} (f, t) \equiv \frac{1}{\rho_{tot} (t)} \cdot \left(\frac{d\rho_{gw}}{d\ln f} \right)     \; ,
\end{equation}
where $\rho_{tot} (t) =3H^2  M^2_{Pl}$ is the total energy density of the Universe; $d\rho_{gw}/d\ln f$ is the spectrum of GW energy density per logarithmic frequency interval. 
Note that $d\rho_{gw}/d\ln f$ is related to $\rho_{gw} (t)$ by 
\begin{equation}
\rho_{gw} (t)=\int^{+\infty}_0 \frac{df}{f} \cdot \left(\frac{d\rho_{gw}}{d\ln f} \right) \; .
\end{equation}
According to the discussion above, the maximum of GW emission occurs
at the time $t_i$ of the domain walls formation. Numerical simulations
of Ref.~\cite{Hiramatsu:2013qaa} then suggest the following fit for
the quantity $d\rho_{gw}/d\ln f$ at its peak\footnote{Note that there is a difference from the analogous expression (33) in Ref.~\cite{Hiramatsu:2013qaa}, where the fractional energy density of gravitational waves $\Omega_{gw}$ is evaluated at much later times, around $t_{dec} \gg t_i$; the time $t_{dec}$ corresponds to the domain walls decay due to explicit (tiny) breaking of $Z_2$-symmetry. 
This difference is due to the fact that the energy density of
gravitational waves $\rho_{gw}$ is constant in the standard case with
a constant tension assumed in Ref.~\cite{Hiramatsu:2013qaa}, and thus
$\Omega_{gw}$ grows until the domain walls decay time $t_{dec}$. On
the contrary, in our scenario, emission of gravitational waves is
efficient only around the time $t_i$, since the tension of domain
walls quickly decreases with time. Note also the different definitions of the Planck mass in our work and in Ref.~\cite{Hiramatsu:2013qaa}.}:
\begin{equation}
\label{logf}
\frac{d\rho_{gw, peak} (t_i)}{d\ln f}  =\frac{\tilde{\epsilon}_{gw} {\cal A}^2 \sigma^2_{wall, i}}{8\pi M^2_{Pl}} \; ,
\end{equation}
which can be equivalently rewritten in terms of the fractional energy
density at the peak:
\begin{equation}
\label{fractiongen}
\Omega_{gw, peak} (t_i) \approx \frac{\lambda \tilde{\epsilon}_{gw} {\cal A}^2 \eta^6_i}{27\pi H^2_i M^4_{Pl}} \; ,
\end{equation}
where we used Eq.~\eqref{surface}. The factors $\tilde{\epsilon}_{gw}$ and ${\cal A}$ (the area parameter) quantify efficiency of GW emission and the scaling property, respectively. 
Including numerical error bars, one has: $\tilde{\epsilon}_{gw}=0.7 \pm 0.4$ and ${\cal A}=0.8 \pm 0.1$\,\cite{Hiramatsu:2013qaa}; both are remarkably close to unity. We use the central values of $\tilde{\epsilon}_{gw}$ and ${\cal A}$ in what follows. 
Before the system settles to the scaling regime, 
the coefficients $\tilde{\epsilon}_{gw}$ and ${\cal A}$ experience a strong time-dependence. In particular, $\tilde{\epsilon}_{gw}$ is very small during this period meaning low GW emission. 
This explains the point made at the end of the previous Section that the moment of settling to the scaling regime is more important for GW phenomenology than the time of domain wall formation.
Substituting Eqs.~\eqref{tension} and~\eqref{hubble} into Eq.~\eqref{fractiongen}, using Eqs.~\eqref{runaway} and~\eqref{initial}, 
we obtain
\begin{equation}
\label{atproduction}
\Omega_{gw, peak} (t_i) \approx \frac{3 \cdot 10^{-9}\cdot N^4}{B \cdot \beta^2} \cdot \left(\frac{100}{g_* (T_i)} \right)^2 \; .
\end{equation} 
Then for the present contribution of GWs one finds 
\begin{equation}
\label{today}
\Omega_{gw, peak} h^2  \approx \frac{4\cdot 10^{-14} \cdot N^4}{B \cdot \beta^2} \cdot \left(\frac{100}{g_* (T_i)} \right)^{7/3} \, ,
\end{equation}
where $h \approx 0.67$ is the reduced Hubble
constant~\cite{Aghanim:2018eyx}. Hereafter, we omit the time argument,
when referring to present day values of GW quantities. To derive
Eq.~\eqref{today} from Eq.~\eqref{atproduction}, we used the relation
\begin{equation}
\label{redshifted}
\Omega_{gw, peak} h^2  \approx 1.34 \cdot 10^{-5}  \cdot \left(\frac{100}{g_* (T_i)} \right)^{1/3} \cdot  \Omega_{gw, peak} (t_i)\; .
\end{equation}
The factor $10^{-5}$ on the r.h.s. is mainly due to the redshift of
$\Omega_{gw} (t)$ during the matter-domination
epoch at $T\lesssim 0.8$\,eV, i.e.,  $\Omega_{gw} (t) \propto 1/a(t)$. 

\begin{figure}[tb!]
  \begin{center}
    \includegraphics[width=\columnwidth,angle=0]{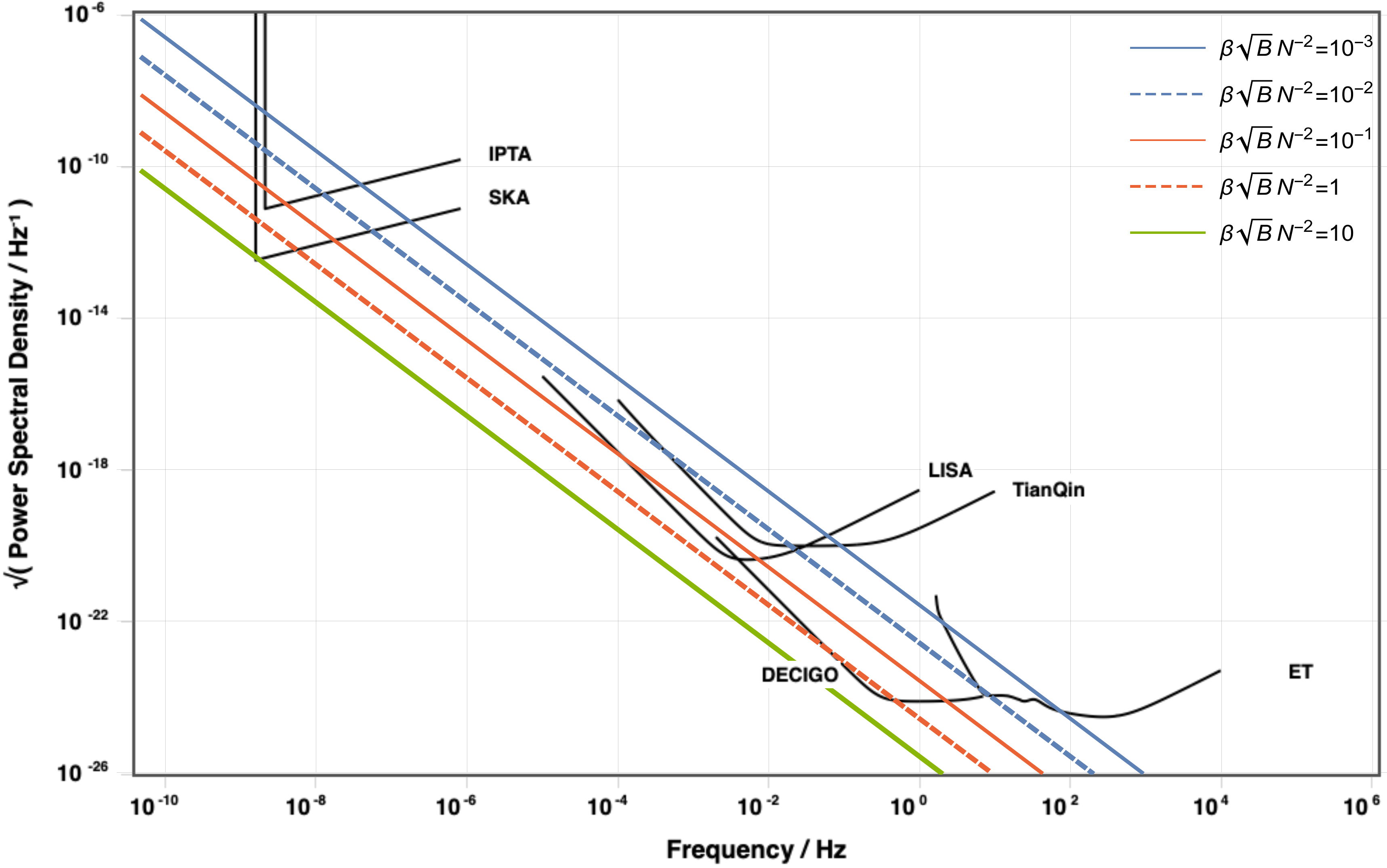}
  \caption{Strain (root power spectral density) $\sqrt{S_h}$
    corresponding to the peak of GW emission from the melting domain walls is shown as a function of characteristic frequency $f_{peak}(t_0)$ for different values of the 
  combined parameter $\beta \sqrt{B}/N^2$. These values have been chosen assuming the range of model constants: $\beta \simeq 1-10$, $B \simeq 1-100$, $N = 4-24$. 
  Note that one point on each colored line corresponds to a particular set of the parameters of the model and it represents the strain at the peak frequency. These colored lines \emph{should not} be confused with the spectrum of GWs shown in Fig.~\ref{spectrum}. Sensitivity curves of some of planned GW detectors shown for comparison are plotted using {\it gwplotter}.}
  \label{gwpeak}
  \end{center}
\end{figure}

Another important characteristic of GWs is their peak frequency
$f_{peak}$, which is estimated as the Hubble rate at the instance of
emission $t_i$, i.e., $f_{peak} (t_i) \simeq
H_i$~\cite{Hiramatsu:2013qaa}. Taking into account the redshift, the peak frequency of GWs reads 
\begin{equation}
\label{fgw}
f_{peak}(t_0)  \simeq \frac{a_i}{ a_0} \cdot H_i  \simeq \left(\frac{g_* (T_i)}{100} \right)^{1/6} \cdot \frac{T_0 T_i}{M_{Pl}} \; ,
\end{equation}
where $T_0 \approx 2.73~\mbox{K}$ is the present day temperature of the Universe. Note that observable frequencies $f_{gw, peak} \lesssim 100~\mbox{Hz}$ correspond to the temperatures $T_i \lesssim 5 \cdot 10^{9}~\mbox{GeV}$. Using the above expression and Eq.~\eqref{initial}, we get  
\begin{equation}
\label{frequency}
f_{peak}(t_0)  \simeq 60~\mbox{Hz} \cdot \left(\frac{N}{B} \right)^{1/2} \cdot \left(\frac{g}{10^{-8}} \right) \cdot \left(\frac{100}{g_* (T_i)} \right)^{1/3} \; .
\end{equation}
One can see that the peak frequency\,\eqref{frequency} and the fractional
energy density of GWs\,\eqref{today} are mainly determined by the
constants $g$ and $\beta$, respectively. Therefore, if the stochastic GW background is detected in the future and attributed to melting domain walls, 
one will be able to determine $g$ and $\lambda$ (recall the relation~\eqref{runaway}) by studying the peak characteristics of the GW background, assuming a reasonable range of $B$ and $N$. 

In Fig.~\ref{gwpeak} we show the characteristic strain of GWs versus the peak frequency for different values of the model parameters $g, \beta, N$ and compare them with sensitivities of Einstein Telescope~\cite{Hild:2010id}, Cosmic Explorer~\cite{Evans:2016mbw}, DECIGO~\cite{Kawamura:2006up}, TianQin~\cite{TianQin:2015yph}, LISA~\cite{Audley:2017drz}, IPTA, and SKA~\cite{Moore:2014eua}, plotted using the publicly available tool {\it gwplotter}~\cite{gwplotter, Moore:2014lga}. The characteristic strain of GWs $\sqrt{S_h}$ is the root of the power spectral density $S_h$ related to the present day fractional energy density of GWs by\footnote{We define the power spectral density as in Refs.~\cite{Moore:2014lga, Romano:2016dpx}.}
\begin{equation}
\Omega_{gw}  H^2_0 =\frac{2\pi^2 f^3}{3} \cdot S_h \; ,
\end{equation}
where $H_0 \approx 67~\mbox{km/(s  Mpc)}$~\cite{Aghanim:2018eyx} is the Hubble constant. 
One can see that altogether the future GW detectors will have a capability to probe the model in the range of portal coupling $10^{-18} \lesssim g \lesssim 10^{-8}$. 
There is a frequency gap between PTAs and LISA, which can be covered by GAIA and THEIA~\cite{Garcia-Bellido:2021zgu}. We confirm one of the main conclusions of Ref.~\cite{Ramazanov:2021eya} that our model is testable in a narrow range 
close to $\beta \simeq 1$, or $\lambda \simeq g^4$, and $N \simeq 10$. Thus, given tiny $g$, we see that very weak self-interactions are phenomenologically most interesting. This feature can be seen as the main 
phenomenological application of melting topological defects: their gravitational radiation can be used to probe (some) field theories in otherwise inaccessible extremely weakly coupled regime, cf. Ref.~\cite{Emond:2021vts}. Compared to our previous work~\cite{Ramazanov:2021eya}, in Fig.~\ref{gwpeak} we also take into account 
the rolling coefficient $B$, which is generically different from unity due to a finite duration of the roll. According to our comment in the end of the previous Section, even in the case 
of a very fast roll to the minimum, it may take some time before the system settles to the scaling regime. For $B \gg 1$ and for the same choice of model parameters, the strain $\sqrt{S_h}$ is reduced 
compared to Ref.~\cite{Ramazanov:2021eya} by the factor $\sqrt{B}$.

\begin{figure}[tb!]
  \begin{center}
    \includegraphics[width=\columnwidth,angle=0]{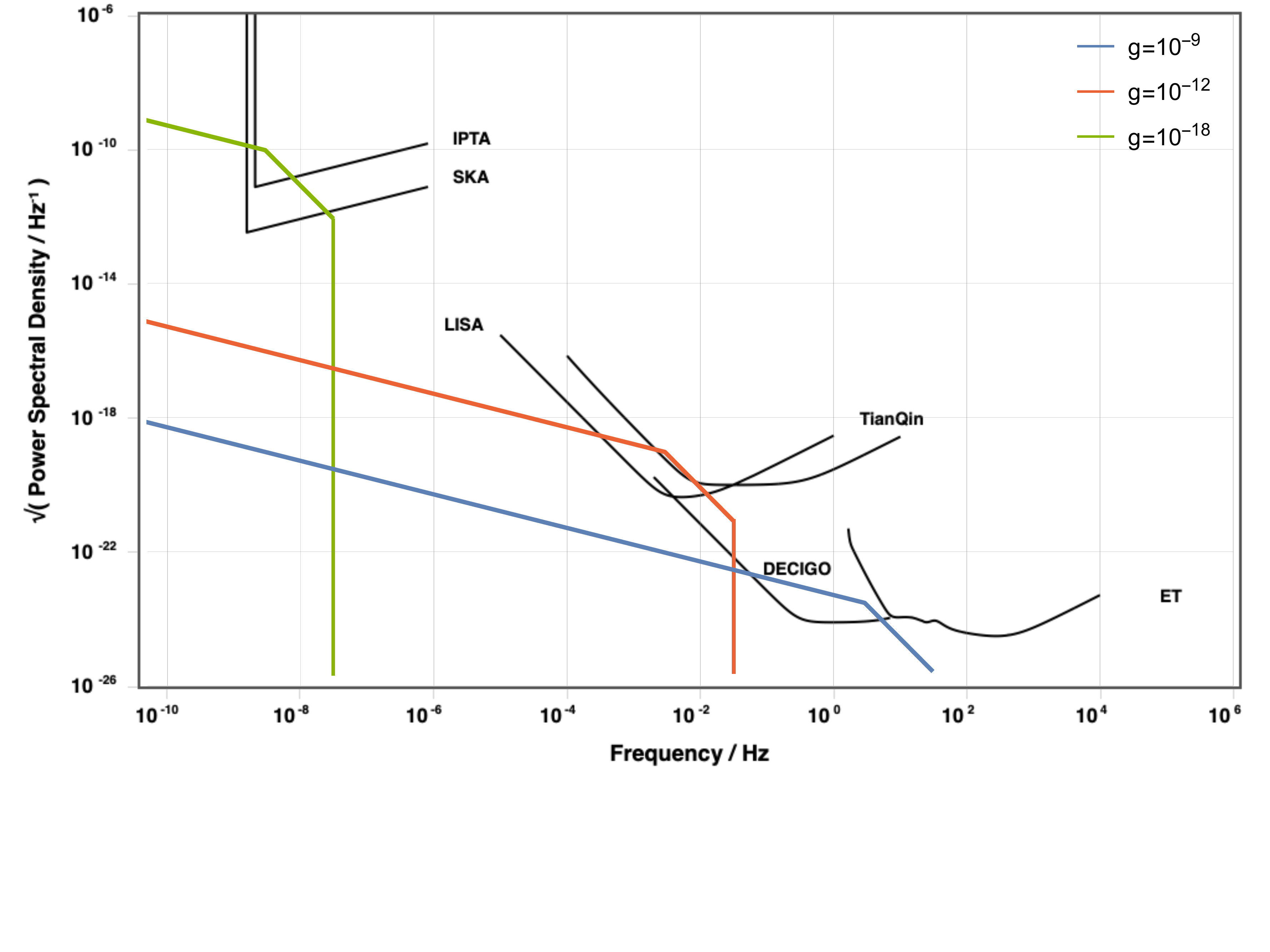}
  \caption{Tentative spectrum of GWs from melting domain walls is shown in terms of strain (root power density) $\sqrt{S_h}$ for different values of constant $g$. Other parameters have been set to $\beta=1$, $N=24$, $B=100$. Sensitivity curves of some of planned GW detectors shown for comparison are plotted using {\it gwplotter}.}
  \label{spectrum}
  \end{center}
\end{figure}

To discriminate GWs emitted by domain walls and other sources, one
needs to know their spectrum. In particular, simulations of
Ref.~\cite{Hiramatsu:2013qaa} reveal the behaviour $\Omega_{gw} (f)
\propto f^3$ at $f<f_{peak}(t_0)$. This numerical result is supported by
considerations of causality. Let us argue that the spectrum of GWs is
different in our case, at least in this low frequency range. In
Ref.~\cite{Hiramatsu:2013qaa}, which assumes a constant tension, the
largest contribution to GWs comes just before the domain walls collapse. In our scenario the situation is reverse: most energetic GWs are emitted close to the
time of wall formation $t_i$. However, GWs continue being emitted at $t>t_i$. This later time emission exactly contributes to the lower frequency part of the GW spectrum\footnote{Here we assume for simplicity that GWs are only emitted at the near-peak frequencies at each moment of time.}. Indeed, the Hubble rate is smaller 
at later times; hence the characteristic frequency is also lower. According to Eq.~\eqref{fgw}, the present day frequency $f$ of GWs emitted at the time, when the temperature of the Universe $T_f$ (where $T_f <T_i$), is defined from 
\begin{equation}
\frac{f}{f_{peak}(t_0)} \simeq \frac{T_f}{T_i} \; .    
\end{equation}
The energy density of this later time GW emission is given by the expression~\eqref{fractiongen}, where one should replace $T_i$ by $T_f$, so that
\begin{equation}
\frac{\Omega_{gw} (f)}{\Omega_{gw, peak} } \simeq \left(\frac{T_f}{T_i} \right)^2\; .    
\end{equation}
Hence, the low frequency tail of GWs can be estimated as 
\begin{equation}
\label{lowspectrum}
\Omega_{gw} (f) \simeq \Omega_{gw, peak} \cdot \left(\frac{f}{f_{peak}(t_0)} \right)^2 \qquad f<f_{peak}(t_0)\;.
\end{equation}
This expression assumes that the late time emission gives the main
contribution to the low frequency part of GW spectrum, and does not
alter the high frequency part of the spectrum. This shows that GWs emitted by melting and constant tension domain walls have clearly distinct spectra.

In the high frequency range $f>f_{peak}(t_0)$ we suppose that the results of Ref.~\cite{Hiramatsu:2013qaa} are applicable\footnote{Naively, one may expect GW spectrum to be different in the high frequency regime for the reasons similar to those presented above. 
Namely, constant tension domain walls emit at the times $t<t_{dec}$, where $t_{dec}$ corresponds to the moment of domain walls collapse, which is also the instance of maximal GW emission. This exactly contributes to frequencies $f>f_{peak}$, because the Hubble rate is larger at earlier times. The analogous contribution is essentially absent in the case of melting domain walls. Note, however, that this contribution decays very fast with the frequency, i.e., $\Omega_{gw} \propto 1/f^4$, and thus can be neglected. }, 
\begin{equation}
\Omega_{gw} (f) \simeq \Omega_{gw, peak} \cdot \left(\frac{f_{peak}(t_0)}{f} \right) \qquad f_{peak}(t_0)<f<f_{cut}(t_0)\; .
\end{equation}
Here $f_{cut}(t_0)$ is the cutoff frequency of GW spectrum roughly defined by the domain wall width, i.e.,
\begin{equation}
f_{cut}(t_0) \simeq (\delta_{w,i})^{-1} \cdot \frac{a_i}{a_0} \simeq f_{peak}(t_0) \sqrt{B} \; .
\end{equation}
As the cutoff frequency $f_{cut}(t_0)$ exceeds $f_{peak}(t_0)$ by a factor $\sqrt{B}$, which is typically not very large, we expect that the spectrum of GWs has a very short tale at high frequencies.
This is again in contrast to the standard 
scenario with constant tension domain walls, where the domain wall thickness is usually much smaller than the horizon size. As a consequence, in the standard scenario one expects an extended high-frequency tale.
This difference can be also used to distinguish 
GW spectrum in our case and in the scenario with constant tension domain walls.

Finally, let us comment on the effect of GWs on BBN. To obtain the contribution of GWs to the energy budget of the Universe at the times of BBN, one multiplies Eq.~\eqref{atproduction} by 
\begin{equation}
\frac{\rho_{tot} (t_i)}{\rho_{tot} (t_{BBN})} \cdot \left(\frac{a_i}{a_{BBN}} \right)^4 \approx \left(\frac{g_* (T_{BBN})}{g_* (T_i)} \right)^{1/3} \; ,
\end{equation}
and gets
\begin{equation}
\label{atBBN}
\Omega_{gw, peak} (t_{BBN})  \approx \frac{ 7 \cdot 10^{-10} \cdot g^{1/3}_{*} (T_{BBN})  \cdot N^4}{B \cdot \beta^2} \cdot \left(\frac{100}{g_* (T_i)} \right)^{7/3} \; .
\end{equation}
Hence, GWs constitute a tiny fraction of the total radiation energy density, unless we deal with an extensively large $N ={\cal O} \left(100 \right)$. Therefore, 
no sizable effect on BBN is expected in our scenario.

\section{Gravitational waves from domain walls formed at preheating}

The picture described in Sections~3 and 4 
assumes that domain walls emerge at
the radiation-dominated epoch. This is a self-consistent scenario, provided that the reheating temperature $T_{reh}$ is larger than the temperature $T_h$ given by Eq.~\eqref{initial}, i.e., $T_{reh}>T_h$.
However, for sufficiently low reheating temperatures and/or small constants $g$, this inequality cannot be fulfilled, and one has
\begin{equation}
\label{condTreh}
T_{reh} \ll  \frac{\sqrt{N}gM_{Pl}}{\sqrt{ g_*(T_{reh})}} \; .
  \end{equation}
In the present Section, we are interested in the situation, when the condition~\eqref{condTreh} is obeyed. In that case, domain walls start to form at the preheating stage. 
The condition~\eqref{condTreh} can be rephrased as a limit on the model constant $g$:
\begin{equation}
g \gg \frac{10 T_{reh}}{N^{1/2} M_{Pl}} \cdot \left(\frac{g_* (T_{reh})}{100} \right)^{1/2} \; .
\end{equation}
For high reheating temperatures (e.g., about the scale of
Grand Unification $\sim 10^{16}$\,GeV or the scale $\sim 10^{14}$\,GeV
where the gauge interactions can
thermalize the plasma of the Standard Model (SM) particles) this estimate implies the peak
frequency of GWs\,\eqref{frequency} well beyond the reach of current and planned experiments. On the other hand, there are well-motivated inflationary models which predict
sufficiently low reheating temperatures. For example, in Starobinsky model\,\cite{Starobinsky:1980te} the
reheating occurs due to the gravitational decay of an inflaton. Assuming that there are only SM degrees of freedom besides an inflaton, the decay rate is sufficiently low, and the resulting reheating temperature is also low, i.e., $T_{reh} \simeq 3 \cdot 10^{9}~\mbox{GeV}$~\cite{Gorbunov:2010bn}.
This corresponds to the marginal value of the constant $g \simeq 10^{-8}$, which is of
interest for the searches of GWs by the Einstein Telescope or Cosmic
Explorer, see Eq.\,\eqref{frequency} and Fig.\,\ref{gwpeak}.

We model the temperature dependence on the scale factor during preheating as a broken powerlaw with positive coefficients $\gamma_1$ and $\gamma_2$: 
\begin{align}
\label{slopes}
T=
T_{max} \cdot \begin{cases}
 \left(\frac{a}{a_{max}} \right)^{\gamma_1}\,, \quad a<a_{max} \\
 \left(\frac{a_{max}}{a} \right)^{\gamma_2}\,, 
\quad a>a_{max} \; .
\end{cases}
\end{align}
Here $T_{max}$ is the maximal temperature of the Universe (which can be higher than the reheating temperature~\cite{Giudice:2000ex}), 
and $a_{max}$ is the scale factor at the moment, when the Universe temperature is maximal. The cosmological evolution during preheating is characterized by the Hubble rate
\begin{equation}
\label{eos}
H^2 \propto \frac{1}{a^{3(1+w)}} \; ,
\end{equation}
where $w$ is the equation of state of the dominant matter. 

For any values of coefficients $\gamma$ and the equation of state $w$, the temperature at the onset of the roll $T_h$ is smaller during preheating compared to the case of radiation-dominated stage for the same choice of model parameters. This follows from the inequality
\begin{equation}
\frac{gN^{1/2}T_h}{\sqrt{12}} \simeq H_h > \sqrt{\frac{\pi^2 g_* (T) }{90 }} \cdot \frac{T_h^2}{M_{Pl}} \; .
\end{equation}
Here we used Eq.~(\ref{beginning}) as well as the assumption of this Section that the rolling of the field $\chi$ starts at preheating, when radiation does not fully determine the Hubble rate. From the above inequality we have
\begin{equation}
\label{less}
T_h < \frac{g N^{1/2}M_{Pl}}{10} \cdot \left(\frac{100}{g_* (T_h)} \right)^{1/2} \; .
\end{equation}
As it follows from Eqs.~\eqref{initial} and~\eqref{surface}, the
tension of domain walls formed at preheating is typically smaller than
the tension of domain walls created at the radiation-dominated stage. 
This conclusion may be altered, if the rolling parameter $B$ is much smaller during preheating than during radiation-domination meaning a shorter time required 
for the field $\chi$ to reach the minimum of its potential.

Note that 
the temperature of the Universe and hence the tension may increase
during preheating after the phase transition. Nevertheless, the maximal temperature $T_{max}$ typically does
not exceed $T_i$ given by Eq.~\eqref{initial}. This follows from the chain of inequalities: 
\begin{equation}
\sqrt{\frac{\pi^2 g_* (T_{max})}{90}} \cdot \frac{T^2_{max}}{M_{Pl}} < H_{max} \lesssim H_h < \frac{g^2 M_{Pl} N}{30} \cdot \left(\frac{100}{g_* (T_h)} \right)^{1/2}  \; ,
\end{equation} 
where the last inequality follows from Eq.~\eqref{beginning} with Eq.~\eqref{less} substituted. We result with the constraint
\begin{equation}
\label{maxlimit}
T_{max} < \frac{gN^{1/2} M_{Pl}}{10} \cdot \left(\frac{100}{g_* (T_h)} \right)^{1/4} \cdot \left(\frac{100}{g_* (T_{max})} \right)^{1/4} \; .
\end{equation}
Comparing the latter with Eq.~\eqref{initial}, we observe that for the same choice of model parameters the maximal temperature is typically lower than the temperature at the onset of the roll $T_h$ in the scenario, where domain walls are formed at radiation-domination ($T_{max}$ may exceed
$T_h$ given by Eq.~\eqref{initial}, if initially there is a small number of degrees of freedom in plasma at the
preheating). Thus, at least for not very large values of the rolling parameter $B$, one expects the maximal temperature $T_{max}$ to be lower than $T_i=T_h/\sqrt{B}$ given by Eq.~\eqref{initial}. As a result, the largest amount of GWs is expected to be
produced by the domain walls formed at the radiation-dominated stage
for the same choice of model parameters. Below we show that it is
indeed the case except the models where the dominant component of energy
density falls faster than radiation at preheating.

Depending on $\gamma_2$ and on the time of domain wall emergence, there are three subcases. We start with 
\begin{equation}
\label{caseone}
\gamma_2 >2/3\,,\qquad \text{and} \qquad t_h< t_{max}\; .
\end{equation}
In this case, the quantity $\rho_{gw} (t) \cdot a^4 (t) \propto T^6 (t) \cdot a^4 (t)$ grows at the times $t<t_{max}$ and then falls at the times $t>t_{max}$. Hence, 
most energetic GWs are radiated, when the Universe temperature reaches $T_{max}$ and thus the present day value of the fractional energy density at peak is given by
\begin{equation}
\Omega_{gw, peak}  =\frac{1}{\rho_{tot, 0}} \cdot \frac{d\rho_{gw, peak} (t_{max})}{d\ln f} \cdot \left(\frac{a_{max}}{a_0} \right)^4 \; .
\end{equation}
 The energy density per logarithmic frequency interval at peak is given by Eq.~\eqref{logf}, where one should replace $i \rightarrow max$: 
\begin{equation}
\frac{d\rho_{gw, peak} (t_{max})}{d\ln f} =\frac{\tilde{\epsilon}_{gw} \cdot {\cal A}^2 \cdot \sigma^2_{wall} (t_{max})}{8\pi M^2_{Pl}} \; .
\end{equation}
Using Eqs.~\eqref{tension} and~\eqref{surface}, we can rewrite the latter as 
\begin{equation}
\label{atmax}
\frac{d\rho_{gw, peak} (t_{max})}{d\ln f} \approx \frac{2 \cdot 10^{-5} \cdot \tilde{\epsilon}_{gw} \cdot {\cal A}^2 \cdot T^6_{max} \cdot N^3}{g^2 \cdot \beta^2 \cdot M^2_{Pl}}  \; .
\end{equation}
It is convenient to split the expression for the fractional energy density of GWs at peak as follows:  
\begin{equation}
\Omega_{gw, peak}  =\frac{1}{\rho_{tot} (t_{reh})} \cdot \frac{d\rho_{gw, peak} (t_{max})}{d\ln f} \cdot \left(\frac{a_{max}}{a_{reh}} \right)^4 \cdot \left(\frac{a_{reh}}{a_0} \right)^4 \cdot \frac{\rho_{tot} (t_{reh}) }{\rho_{tot,0} } \; .
\end{equation}
Here we substitute Eq.~\eqref{atmax}, use $\rho_{tot} (t_{reh})=\pi^2 g_* (T_{reh})T^4_{reh}/30$ and (cf. Eq.~\eqref{redshifted})
\begin{equation}
 \left(\frac{a_{reh}}{a_0} \right)^4 \cdot \frac{\rho_{tot} (t_{reh})}{\rho_{tot,0}} \approx \frac{1.34}{h^2} \cdot 10^{-5} \cdot \left( \frac{100}{g_* (T_{reh})} \right)^{1/3} \; .
\end{equation}
We also express the ratio of scale factors through the ratio of temperatures using Eq.~\eqref{slopes}, i.e., $a_{reh}/a_{max} \simeq (T_{max}/T_{reh})^{1/\gamma_2}$, substitute the values $\tilde{\epsilon}_{gw} \approx 0.7$ and ${\cal A} \approx 0.8$~\cite{Hiramatsu:2013qaa}, and obtain finally
\begin{equation}
\label{energymaxp}
\Omega_{gw, peak} h^2 \simeq \frac{4 \cdot 10^{-12}  \cdot T^2_{max} \cdot N^3}{g^2 \cdot \beta^2 \cdot M^2_{Pl}} \cdot \left(\frac{T_{max}}{T_{reh}} \right)^{4-\frac{4}{\gamma_2}} \cdot \left(\frac{100}{g_* (T_{reh})}  \right)^{4/3} \; .
\end{equation}
Comparing the latter with Eq.~\eqref{today} and using
Eq.~\eqref{maxlimit}, we see that for $\gamma_2 < 1$ the peak
energy density of GWs is typically smaller in the case~\eqref{caseone}
than in the case of domain walls formed at the radiation-dominated
stage (except the cases of large $B$ at radiation-domination and small number of initial degrees of freedom at preheating mentioned above). In the case $\gamma_2=1$, which corresponds to
effective radiation domination (e.g. due to inharmonic quartic
oscillator $\varphi^4$), Eq.\,\eqref{energymaxp} yields the same result as at the true hot stage discussed in
previous Section, provided that $T_{max}$ saturates the upper bound in Eq.~\eqref{maxlimit}.

The same conclusion about GWs is not necessarily true for $\gamma_2>1$, which suggests a
stiff equation of state  $w>1/3$ during preheating\footnote{Indeed, only for $w>1/3$ the energy density of produced radiation $\rho_{rad} 
\propto T^4 \propto 1/a^{4\gamma_2}$ can redshift slower than the dominant energy density of the Universe $\rho_{tot} \propto 1/a^{3(1+w)}$.}. As a result, the energy density of GWs redshifting as $1/a^4$ grows relative to the energy density of the dominant matter redshifting as $1/a^{3(1+w)}$. This leads to the amplification of GWs, which counteracts the suppression following from Eq.~\eqref{maxlimit}. 
Note also that Eq.~\eqref{energymaxp} does not involve the suppressing coefficient $B$, which takes into account the time of the 
roll. This is true as long as the preheating stage is substantially long, so that there is enough time for domain walls to be formed and enter the scaling regime, before the Universe temperature reaches $T_{max}$. 

The second subcase corresponds to
\begin{equation}
\gamma_2<2/3  \; ,
\end{equation}
for which the quantity $\rho_{gw} \cdot a^4 (t)$ grows until the Universe gets reheated, so that the maximum of GWs is emitted, when the temperature reaches $T_{reh}$. The present day peak fractional energy density of GWs is obtained 
from Eq.~\eqref{energymaxp} by formally identifying $T_{max}$ and $T_{reh}$:
\begin{equation}
\label{gwreheating}
\Omega_{gw, peak} h^2 \simeq \frac{4 \cdot 10^{-12}  \cdot T^2_{reh} \cdot N^3}{g^2 \cdot \beta^2 \cdot M^2_{Pl}} \cdot \left(\frac{100}{g_* (T_{reh})} \right)^{4/3} \; .
\end{equation}
As it follows from Eq.~\eqref{maxlimit}, where one should replace $T_{max}$ with $T_{reh}$, again less GWs are expected in this scenario compared to the case of domain walls formed entirely at radiation-dominated stage. This reduction of GW emission is partially compensated by the fact that the 
coefficient $B$ is absent. Note that the peak frequency of GWs is tightly pinned to the reheating temperature of the Universe: 
\begin{equation}
\label{frreheating}
f_{peak}(t_0)  \simeq \left(\frac{g_* (T_{reh})}{100} \right)^{1/6} \cdot \frac{T_0 T_{reh}}{ M_{Pl}} \; .
\end{equation}
Therefore, in this scenario measuring the peak frequency of GWs, one can define the reheating temperature of the Universe.

Finally, in the situation, when 
\begin{equation}
\gamma_2>2/3 \qquad t_h >t_{max} \; ,
\end{equation}
the maximum of GWs is reached at the time of domain wall formation $t_i$, which corresponds in this case to the intermediate temperature between the reheating and maximal one: 
\begin{equation}
\label{intermone}
\Omega_{gw, peak} h^2 \simeq  \frac{4 \cdot 10^{-12}  \cdot T^2_{i} \cdot N^3}{g^2 \cdot \beta^2 \cdot M^2_{Pl}} \cdot \left(\frac{a_{reh}}{a_i} \right)^{4\gamma_2-4} \cdot \left(\frac{100}{g_* (T_{reh})} \right)^{4/3}\; .
\end{equation}
In the above expression one can eliminate the dependence on $T_i$ and $a_i$ and express 
the energy density of GWs in terms of model constants and parameters,
which characterize evolution of Universe at preheating, i.e.,
$T_{reh}$, $T_{max}$, $w$, and $\gamma_2$. The resulting expression is
not very illuminating and we present  
it in Appendix~A. 

The discussion above assumes a sufficiently slow preheating, so that
the roll to the minimum occurs entirely at one or another slope of
Eq.~\eqref{slopes}. Let us comment on the case of an almost instantaneous
reheating. In that case, the field $\chi$ instantly acquires the thermal mass. If the latter is smaller than the Hubble rate, then we end up with the storyline of previous Sections: domain walls form and evolve entirely during the radiation-dominated stage. It is possible, however, that the thermal mass is larger than the Hubble rate, i.e., $\sqrt{\lambda} \eta_{reh} =\zeta H_{reh}$, where $\zeta>1$ is the constant factor. It is clear that the field $\chi$ starts to roll immediately, as the thermal mass appears. The temperature at the start of the roll is set by the reheating temperature:
\begin{equation}
\label{reheating}
T_h = T_{reh} \simeq \frac{N^{1/2} gM_{Pl}}{10\zeta} \cdot \left(\frac{100}{g_* (T_{reh})} \right)^{1/2}\; .
\end{equation}
Due to the large thermal mass relative to the Hubble parameter at the onset of the roll, it takes less time for the field $\chi$ to reach the minimum resulting in smaller values of the coefficient $B$ compared to Eq.~\eqref{A}:
\begin{equation}
B=\mbox{max} \left[\frac{1}{\zeta^2} \ln^2 \left(\frac{2\eta_h}{\delta \chi_h} \right), 1 \right]\; .    
\end{equation}
For sufficiently large $\zeta$, one has $B \simeq 1$ and the set of predictions regarding GWs is given by Eqs.~\eqref{gwreheating} and~\eqref{frreheating}.

In the end of this Section, we would like to stress an important point regarding GW signals at
preheating. While typically the amount of produced GWs is lower than at radiation-domination, the peak frequency is different. Hence, for the same values of the model parameters, 
cosmological models with low reheating temperatures may have better chances to be probed with future GW detectors.

\section{Late time evolution: Dark Matter}

So far we have been dealing with a scale-invariant model~\eqref{basic}. In this Section we introduce a slight breaking of scale-invariance by adding masses to the fields $\chi$ and $\phi$, which we assume to be much smaller than $gT_i$ and $T_i$, respectively, where $T_i$ is the temperature of domain wall formation. In particular, we extend the Lagrangian~\eqref{basic} of the field $\chi$ by
\begin{equation}
{\cal L}_{mass}=-\frac{M^2 \chi^2}{2} \; .
\end{equation}
We assume that the mass squared $M^2$ is positive 
\begin{equation}
\label{masschiplus}
M^2>0 \; .
\end{equation}
An alternative option with the negative mass squared will be
considered in the next Section. We mostly assume that the mass term
for the field $\phi$ is also non-tachyonic:
\begin{equation}
\label{massphiplus}
m^2_{\phi}>0 \; .
\end{equation}
Occasionally we also comment on the case $m^2_{\phi}<0$, which leads
to the spontaneous symmetry breaking in the field $\phi$ sector at low temperatures and occurs, e.g., 
in the case of Higgs portal, when $\phi$ is the SM Higgs doublet.
Once Eq.~\eqref{masschiplus} and/or Eq.~\eqref{massphiplus} are
fulfilled, the inverse phase transition inevitably happens, and the domain walls are annihilated, because the expectation value of the field $\chi$ given by 
\begin{equation}
\label{min}
\langle \chi \rangle=\sqrt{\eta^2 (T)-\frac{M^2}{\lambda}} \; ,
\end{equation}
eventually sets to zero. Recall that $\eta (T)$ is given by Eq.~\eqref{tension} and it has the meaning of the expectation value of the field $\chi$ at high temperatures. 
  
  The temperature of the inverse phase 
transition depends on the ratio of the masses $M$ and $m_{\phi}$. For substantially light particles $\phi$, i.e.,
\begin{equation}
\label{l}
 m_{\phi} \lesssim \sqrt{\frac{12}{N}} \cdot \frac{M}{g} \; ,
\end{equation}
one can use the expression~\eqref{tension} all the way down to the moment of the inverse phase transition. Thus, the temperature of the symmetry restoration is given by
\begin{equation}
\label{light}
T_{sym} \simeq \sqrt{\frac{12}{N}} \cdot \frac{M}{g} \; .
\end{equation}
Instead, for substantially heavy particles $\phi$, i.e., 
\begin{equation}
\label{h}
m_{\phi} \gtrsim \sqrt{\frac{12}{N}} \cdot \frac{M}{g} \; ,
\end{equation}
the symmetry restoration occurs (for the same $M$ and $g$) at higher temperatures,
\begin{equation}
\label{heavy}
T_{sym} \simeq m_{\phi}  \; .
\end{equation}
The reason is that the quantity $\eta (T)$ rapidly approaches zero due
to the Boltzmann suppressing factor, $\eta^2 (T) \propto
e^{-\frac{m_{\phi}}{T}}$, as soon as the particles $\phi$ become
non-relativistic, but remain in equilibrium with relativistic
particles in the primordial plasma. Note that the temperature $T$ associated with the 
particles $\phi$ can be different from the temperature of the
Universe, which is the case, when the particles $\phi$ had decoupled from the SM species at some point in 
preceding cosmological history. In this situation, the issues related to the lifetime of particles $\phi$ and embedding the latter into 
a proper particle physics framework become important. Therefore, we do not consider the scenario with out-of-equilibrium particles $\phi$, unless otherwise specified.

Promoting $\chi$ to a massive field, one can consider it for the role
of DM.  Note that for tiny $g$ we are interested in, the standard
freeze-out mechanism of DM production is not applicable, because
$\chi$ is never in equilibrium with primordial plasma. The most common way of generating feebly coupled DM is by the freeze-in, i.e.,
through the out-of-equilibirum scatterings of (beyond) the SM
particles~\cite{McDonald:2001vt, Hall:2009bx}.  Still, this mechanism
involves couplings $g \simeq 10^{-6}$~\cite{Chu:2011be, Yaguna:2011qn,
  Lebedev:2019ton, Bringmann:2021sth} (see the discussion in Subsection 6.3), which
implies too high peak frequency of GWs\,\eqref{frequency}, well above
the range to be explored with present projects, see
Fig.\,\ref{gwpeak}.  Another mechanism relying on the inverse phase
transition and operating for extremely weak couplings $g$, has been
discussed in Ref.~\cite{Ramazanov:2021eya} and we review it in
Section\,\ref{subsec:inversed-2order}. Before doing that, in Section\,\ref{subsec:2order} we put forward another mechanism, where DM is
sourced by oscillations following relaxation to the minimum of
spontaneously broken phase.

\subsection{Dark Matter from the second order phase transition at
  radiation-domination}
\label{subsec:2order}

During the second order phase transition, the field $\chi$ rolls from $\chi=0$ to the minimum of effective potential~\eqref{spontaneous} and develops a large velocity. As a result, the field 
$\chi$ does not settle in the minimum, but rather starts to oscillate around it according to Eq.~\eqref{flat}, where one should substitute $\tilde{M}_{eff} =\sqrt{2\lambda} \tilde{\eta}=\mbox{const}$: 
\begin{equation}
\label{osc}
\delta \tilde{\chi} \equiv \tilde{\chi} -\tilde{\eta} =\delta \tilde{\chi_i} \cdot \sin \left(\int^{\tau}_{\tau_i} d \tau' \tilde{M}_{eff} (\tau') \right) \; .
\end{equation}
(For concreteness, we consider oscillations around the positive minimum value.) Equating the potential energy density of the field $\tilde{\chi}$ at the onset of the roll, $\lambda \tilde{\eta}^4/4$, to its kinetic energy density 
on the bottom of the effective potential, $\tilde{M}^2_{eff} \delta \tilde{\chi_i}^2/2$, and returning to the original field variable, we get the amplitude $\delta \chi_i$ of oscillations:
\begin{equation}
\label{amplitude}
\delta \chi_i \simeq \frac{\eta_i}{2} \; .
\end{equation}
The oscillations of the field $\chi$ contribute to DM at sufficiently late times. Naively, one may expect these oscillations to decay into the particles $\phi$. However, the latter have 
thermal masses $m_{\phi} (T) \gtrsim \sqrt{\lambda_{\phi}} T$ much
larger than $M_{eff} =\sqrt{2\lambda} \eta$, at least in the region of
model parameters interesting for GW observations, and therefore the
perturbative decay $\chi\to\phi$ is 
prohibited. The non-perturbative decay, e.g., through the narrow parametric resonance is not efficient, because there is no Bose--Einstein amplification: particles $\phi$ produced by oscillations of the field $\chi$ quickly thermalize with the hot plasma. We conclude that the field $\chi$ remains stable in the spontaneously broken 
phase, unless it has interactions with other substantially light species. The latter option will be considered in the next Subsection. 

\begin{figure}[tb!]
  \begin{center}
    \includegraphics[width=0.32\columnwidth,angle=0]{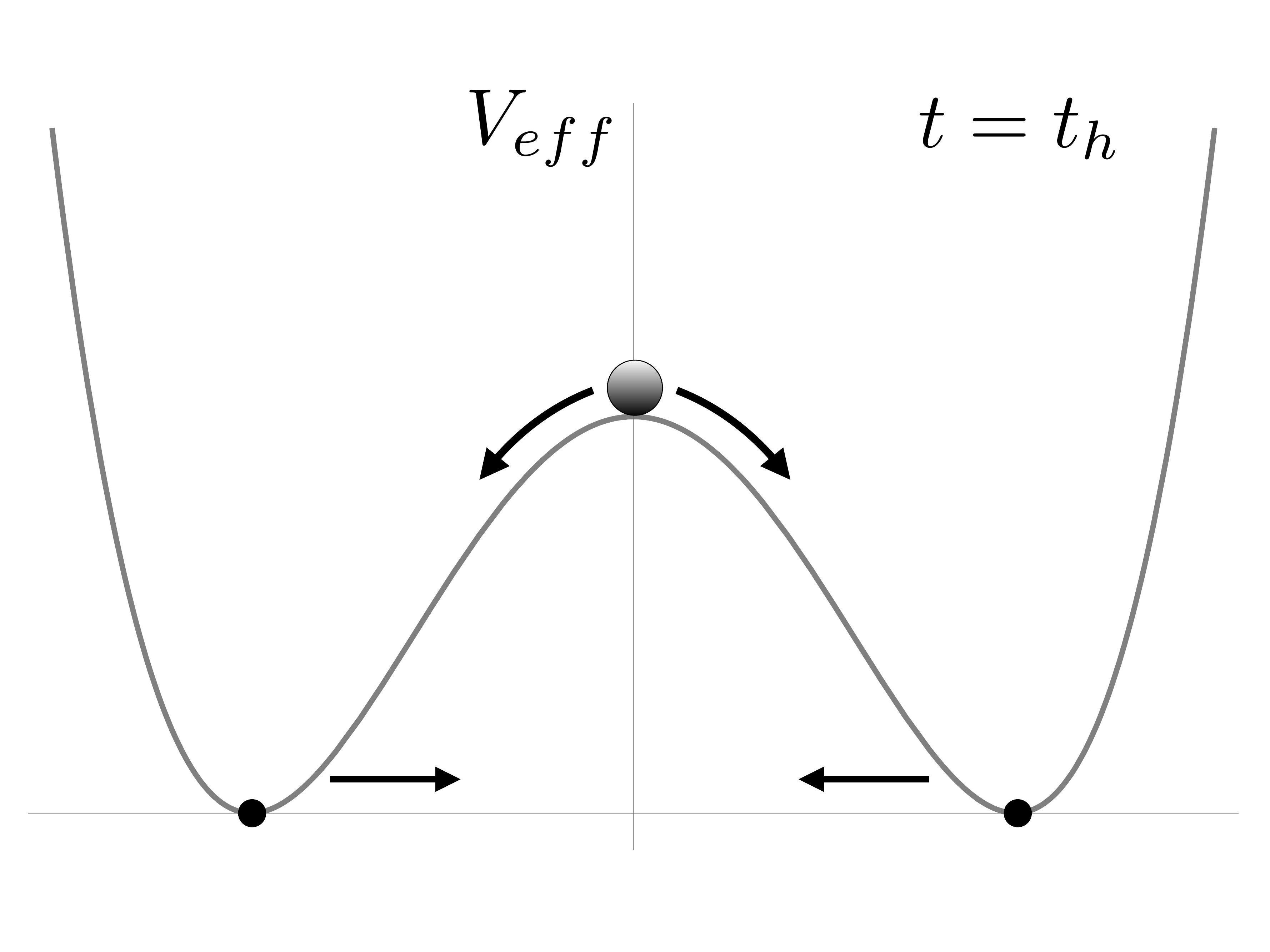}
        \includegraphics[width=0.32\columnwidth, angle=0]{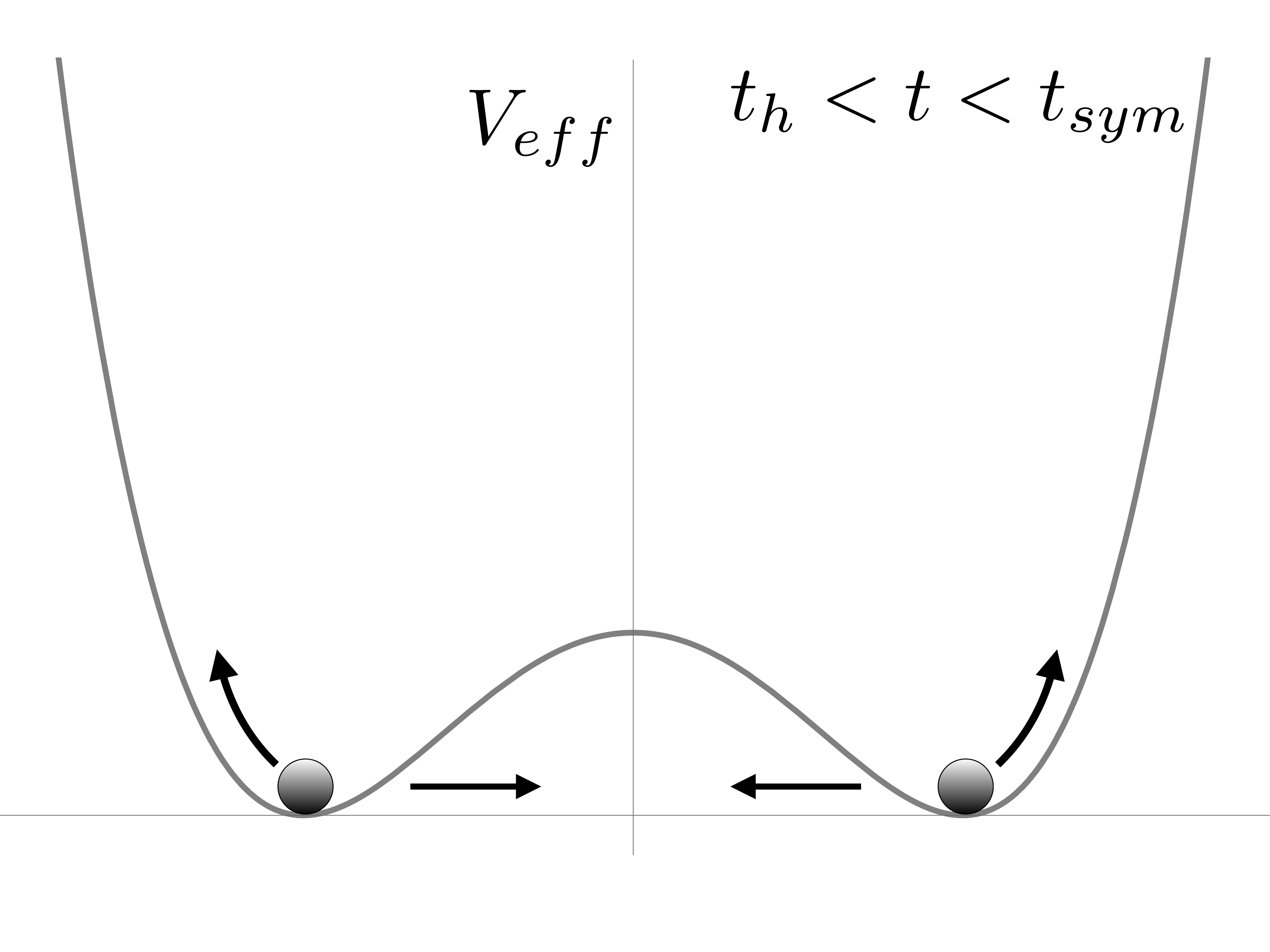}
          \includegraphics[width=0.32\columnwidth, angle=0]{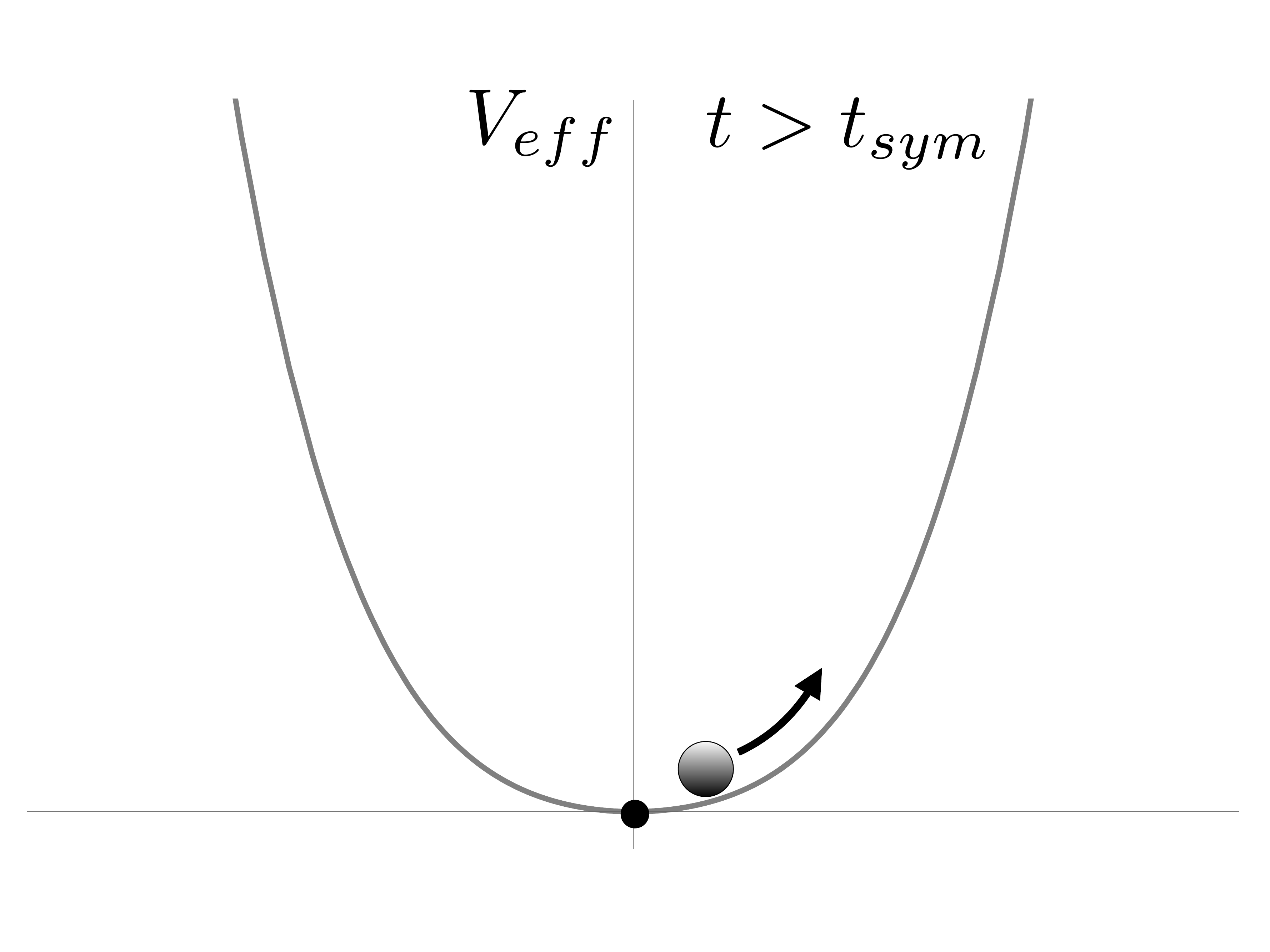}
  \caption{DM production at the second order phase transition. The field $\chi$ starts to roll to the minimum of the broken phase at the time $t_h$ (left figure) and then oscillates around the non-trivial minima~\eqref{min} 
  at the times $t_h<t<t_{sym}$ (figure in the middle). These oscillations continue, now around the trivial minimum at $\chi=0$, after the symmetry is restored, $t>t_{sym}$ (right figure). The energy density contained in the field $\chi$ oscillations 
  has a radiation-like behaviour at early times, which turns into the matter-like behaviour, as soon as the mass term of the field $\chi$ comes into play. }\label{dpt}
  \end{center}
\end{figure}

Initially, the energy density of oscillations is estimated by
\begin{equation}
\rho_{\chi, i} \simeq \frac{M^2_{eff, i} \cdot \delta \chi^2_i}{2}  \simeq \frac{N^2 T^4_i}{600\beta}\; .
\end{equation}
Before the bare mass term of the field $\chi$ comes into play at the time $t_{dm}$, the amplitude of oscillations $\delta \chi$ decreases as
\begin{equation}
\label{radlike}
\delta \chi (t) \simeq \delta \chi_i \cdot \frac{a_i}{a(t)} \; .
\end{equation}
Thus, at $t \lesssim t_{dm}$ the energy density $\rho_{\chi}$ of the field $\chi$ oscillations redshifts as radiation. At $t \simeq t_{dm}$, when evolution of the field $\chi$ becomes dominated 
by the bare mass term, conventional DM phase starts with the amplitude of oscillations decreasing as $\delta \chi \propto 1/a^{3/2}$. The energy density of this DM is given by
\begin{equation}
  \label{DM-e}
\rho_{\chi} (t>t_{dm}) \approx \rho_{\chi, i} \cdot \left(\frac{a_{i}}{a_{dm}} \right)^4 \cdot \left(\frac{a_{dm}}{a(t)} \right)^3 \simeq \frac{N^2 T^4_{dm}}{600\beta} \cdot \left(\frac{g_* (T_{dm})}{g_* (T_i)} \right)^{4/3} \cdot \left(\frac{a_{dm}}{a(t)} \right)^3 \; .
\end{equation}
Assuming that the field $\chi$ makes all of DM in the Universe, we can define $T_{dm}$ using the condition $\rho_{\chi} (t_{eq}) \approx \rho_{rad} (t_{eq})$, which corresponds to the matter-radiation equality at the temperature $T_{eq} \approx 0.8~\mbox{eV}$: 
\begin{equation}
\label{directtemperature}
T_{{dm}} \simeq 50~\mbox{keV} \cdot \left(\frac{g_* (T_i)}{100} \right)^{4/3} \cdot \left(\frac{3.4}{g_* (T_{dm})} \right)^{1/3} \cdot \frac{\beta}{N^2}  \; .
\end{equation}
Note that the temperature $T_{dm}$ is independent of the mass $m_{\phi}$, which only affects the relation between $T_{dm}$ and $T_{sym}$. For sufficiently heavy particles $\phi$ described by Eq.~\eqref{h}, the symmetry is restored 
relatively early, when the amplitude of the field $\chi$ is still large enough. In that case, evolution of the field $\chi$ effectively continues in the quartic potential $\sim \lambda \chi^4$ until it drops to $\lambda \chi^4 \sim M^2 \chi^2$. Consequently, Eq.~\eqref{radlike} holds for some time after the symmetry restoration, so that $T_{sym} \gg T_{dm}$. Otherwise, if the particles $\phi$ are light, see Eq.~\eqref{l}, the symmetry restoration is due to the bare mass of the field $\chi$, so that $T_{sym} \simeq T_{dm}$. Both scenarios are phenomenologically indistinguishable. 
Hence, for this DM production scenario it is irrelevant, if the field $\phi$ is heavy of light.

Expressing the ratio $\beta/N^2$ in terms of the temperature $T_{dm}$
via Eq.\,\eqref{directtemperature} and substituting into
Eq.~\eqref{today}, we obtain the fractional energy density of GWs at
the peak frequency in the scenario of interest:
\begin{equation}
\label{directgw}
\Omega_{gw, peak}  h^2 \simeq \frac{10^{-12}}{B} \cdot \left(\frac{g_* (T_i)}{100} \right)^{1/3} \cdot \left(\frac{3.4}{g_* (T_{dm})} \right)^{2/3} \cdot \left(\frac{10~\mbox{keV}}{T_{dm}} \right)^2\; .
\end{equation}
We see that for the GWs to be in the observable range of frequencies
the temperature should be sufficiently low $T_{dm} \lesssim 1-10~\mbox{keV}$ depending on the value of $B$.

Let us discuss the lower bound on the temperature $T_{dm}$, which will
give us the upper bound on the amount of GWs produced in this scenario. 
A conservative bound follows from the fact that oscillations of the field $\chi$ contribute to dark radiation at the temperatures $T \gtrsim T_{dm}$. 
Assuming that oscillations of the field $\chi$ are the only source of dark radiation, we write $\rho_{\chi}  (T_{dm}) \simeq \rho_{dr} (T_{dm}) \simeq \rho_{dm} (T_{dm})$. Using this and the condition $\rho_{dm} (T_{eq}) \simeq \rho_{rad} (T_{eq})$ at the matter-radiation equality, we get 
$\rho_{dm} (T_{dm}) \simeq \rho_{dr} (T_{dm})\simeq \rho_{rad}
(T_{dm}) \cdot \left(T_{eq}/T_{dm} \right)$. For $T_{dm}$ interesting for GW phenomenology, $g_*$ does not change: there are only photons in the hot plasma. Since both ordinary and dark radiation
dependence on the scale factor in the same way, one has at BBN $\rho_{dr} (T_{BBN}) \simeq \rho_{rad} (T_{BBN}) \cdot \left(T_{eq}/T_{dm} \right)$. 
Consequently, the ratio of dark radiation to the ambient energy density at BBN, $F_{dr} \equiv \rho_{dr} (T_{BBN})/\rho_{rad} (T_{BBN})$, is given by 
\begin{equation}
\label{ratio}
F_{dr} \simeq \frac{T_{eq}}{T_{dm}} \; .
\end{equation}
This ratio is commonly expressed via the departure from the effective number 
of neutrino species $\Delta N_{\nu} \equiv N_{\nu}-N_{\nu, SM}$, where $N_{\nu, SM} \approx 3.046$ is the SM value:
\begin{equation}
F_{dr}=\frac{7}{4} \cdot \left(\frac{4}{11} \right)^{4/3} \cdot \frac{\Delta N_{\nu}}{g_* (T_{BBN})} \; .
\end{equation}
Planck data strongly constrained $N_{\nu}$ at BBN: $N_{\nu}=2.99 \pm 0.17$ at $68\%$~\mbox{CL}~\cite{Aghanim:2018eyx} giving $\Delta N_{\nu} \lesssim 0.11$ and hence $F_{dr} \lesssim 0.01$; we used the value $g_* (T_{BBN}) \approx 3.4$ meaning the BBN temperature $T_{BBN} \approx 70~\mbox{keV}$. Now, using Eq.~\eqref{ratio} one gets the bound on the 
temperature $T_{dm}$: 
\begin{equation}
\label{lowerbbn}
T_{dm} \gtrsim 70~\mbox{eV} \; ,
\end{equation}
which implies the upper bound on the fractional energy density of GWs~\eqref{directgw}: 
\begin{equation}
\Omega_{gw, peak} h^2  \lesssim \frac{2 \cdot 10^{-8}}{B} \cdot \left(\frac{g_* (T_i)}{100} \right)^{1/3} \cdot \left(\frac{3.4}{g_* (T_{dm})} \right)^{2/3}  \; .
\end{equation}
This should be viewed as a rather conservative bound, which may be strengthened by the study of DM perturbations and large scale structure formation in our scenario. For example, in the case of sterile neutrino DM the lower bound on the relevant temperature is in the ballpark of $T_{dm} \simeq 1-10~\mbox{keV}$~\cite{Yeche:2017upn}. The latter is not applicable to our case, as it follows from the free streaming of relativistic particles. Nevertheless, we expect a prominent effect on the large scale structure 
in the range of model parameters, which corresponds to most energetic GWs. 

 We relate the mass $M$ to the temperature $T_{dm}$ by making use of Eq.~\eqref{DM-e}, 
 \begin{equation}
   \label{e1}
\frac{M^2 \delta \chi^2_{dm}}{2} \simeq \frac{N^2 T^4_{dm}}{600 \beta} \cdot \left(\frac{g_* (T_{dm})}{g_* (T_i)} \right)^{4/3} \; . 
\end{equation}
The amplitude of DM oscillations $\delta \chi_{dm}$ is given by 
\begin{equation}
  \label{e2}
\delta \chi_{dm} \simeq \delta \chi_i \cdot \left(\frac{a_i}{a_{dm}} \right) \simeq \frac{\eta_i}{2} \cdot \left(\frac{T_{dm}}{T_i} \right) \cdot \left(\frac{g_* (T_{dm})}{g_* (T_{i})} \right)^{1/3} \simeq \frac{N^{1/2}T_{dm}}{4g \sqrt{3\beta}} \cdot 
\left(\frac{g_* (T_{dm})}{g_* (T_{i})} \right)^{1/3} \; .
\end{equation}
From the latter two expressions \eqref{e1} and \eqref{e2} we get
\begin{equation}
\label{massdirect}
M \simeq 1.3 \cdot 10^{-5}~\mbox{eV} \cdot N^{1/2} \cdot \left(\frac{g}{10^{-8}} \right) \cdot \left(\frac{T_{dm}}{10~\mbox{keV}} \right) \cdot \left(\frac{g_* (T_{dm})}{3.4} \right)^{1/3} \cdot \left(\frac{100}{g_* (T_i)} \right)^{1/3} \; .
\end{equation}
Thus, in the range interesting for GW measurements we deal with very light DM having masses typical, e.g., for QCD axions.

At the end of this Subsection, let us briefly comment on the scenario with the the Higgs field playing the role of $\phi$, which corresponds to the choice $m^2_{\phi}<0$. In this case, the field $\chi$ acquires 
a tachyonic mass due to the non-zero expectation value of the Higgs field $v_{SM} \approx 246~\mbox{GeV}$. As we want the overall bare mass squared of the field $\chi$ to be positive, the inequality  $M \gtrsim g v_{SM}$ must be fulfilled. Using Eq.~\eqref{massdirect}, we obtain:
\begin{equation}
T_{dm} \gtrsim \frac{2~\mbox{TeV}}{\sqrt{N}} \cdot \left(\frac{g_* (T_i)}{100} \right)^{1/3} \cdot \left(\frac{3.4}{g_* (T_i)} \right)^{1/3}  \; .
\end{equation}
For these large temperatures $T_{dm}$, the present day energy density
of GWs is very small $\Omega_{gw} h^2 \lesssim 0.3 \cdot 10^{-28}/B$ according to
Eq.~\eqref{directgw}.

\subsection{Dark Matter production at inverse phase transition}
\label{subsec:inversed-2order}

Here we consider the situation, where oscillations of the field $\chi$ following from the direct second order phase transition quickly relax to zero. 
Consequently, the field $\chi$ just tracks the minimum of the broken phase:
\begin{equation}
\label{onset}
\chi  \approx \langle \chi  \rangle  \qquad t_i \ll t \lesssim t_{sym} \; .
\end{equation}
Such a relaxation is possible, provided that the field $\chi$ transfers
its energy to other species, e.g., into some light particles. This
energy loss can occur in the narrow parametric resonance regime and is
analogous to that of an inflaton at preheating~\cite{Kofman:1997yn,
  Greene:1997fu}.
We discuss one example of such a decay in Appendix B.

Once the mass of the field $\chi$ is included, the expectation value $\langle \chi \rangle$ is given by Eq.~\eqref{min}. It is crucial that the derivative $d\langle \chi \rangle/dt$ explodes at the time $t_{sym}$, when $\langle \chi \rangle$ is zero. This grossly non-adiabatic character of the inverse phase transition triggers oscillations of the field $\chi$, which feed into DM energy density~\cite{Ramazanov:2021eya}\footnote{Other implementations of the inverse phase transition for DM production can be found in Ref.~\cite{Baker:2019ndr}. In Refs.~\cite{Dodelson:1989ii, Dodelson:1989cq} the inverse phase transition is used to generate the baryon asymmetry of the Universe.}. This mechanism of DM production is universal: in Refs.~\cite{Babichev:2020xeg} and~\cite{Ramazanov:2020ajq} it has been discussed for the DM field coupled to the Ricci scalar and primordial magnetic fields, respectively.

Let us define the amplitude of the field $\chi$ and its energy density
at the onset of oscillations in terms of the model parameters. In the
discussion below we closely follow Ref.\,\cite{Babichev:2020xeg}. 
For this purpose, we consider the effective mass squared of the field $\chi$:
\begin{equation}
M^2_{eff} \equiv \frac{\partial^2 V_{eff}}{\partial \chi^2} =3\lambda \chi^2 -\lambda \eta^2 (T) +M^2 \; .
\end{equation}
In the broken phase, the effective mass squared is given by
\begin{equation}
\label{invacuum}
M^2_{eff} \approx 2 \lambda \langle \chi \rangle ^2  \; .
\end{equation}
At the times $t \ll t_{sym}$, the quantity $M_{eff}$ changes very slowly with time:
\begin{equation}
\label{ineq}
\left| \frac{\dot{M}_{eff}}{M^2_{eff}} \right| \ll 1 \; .
\end{equation}
That behaviour corresponds to the situation, when the field $\chi$ tracks its minimum, $\chi =\langle \chi \rangle$. The tracking
stops at the time $t_*$, when the quantity of Eq.~\eqref{ineq}
approaches equality, which implies 
\begin{equation}
\label{violation}
\frac{|\dot{\chi}_* |}{\chi^2_*} \simeq \sqrt{2\lambda} \; ,
\end{equation}
where we used Eqs.~\eqref{onset} and~\eqref{invacuum}. We obtain the time derivative $|\dot{\chi}_*|$ from Eqs.~\eqref{min} and~\eqref{onset}: 
\begin{equation}
|\dot{\chi}_*| \simeq \frac{\kappa H_* \eta^2 (T_*)}{|\chi_*|} \simeq \frac{\kappa M^2 H_*}{\lambda |\chi_*|} \; .
\end{equation}
Here we expressed $\eta (T_*)$ in terms of $M$ and $\lambda$, i.e., $\eta (T_*) \approx M/\sqrt{\lambda}$ assuming $T_* \approx T_{sym}$. The coefficient $\kappa$ is defined as 
\begin{equation}
\kappa \equiv \left|\frac{\dot{\eta}}{H \cdot \eta} \right|_{sym} \approx  \left|\frac{\dot{\eta}}{H \cdot \eta} \right|_{*}\; .
\end{equation}
 We get
\begin{equation}
\label{init}
\chi_* \simeq \frac{\kappa^{1/3} (M^2 H_*)^{1/3}}{2^{1/6}\sqrt{\lambda}} \; .
\end{equation}
The amplitude $\chi_*$ defines the energy density of DM comprised of the field $\chi$ in the standard fashion:
\begin{equation}
\label{energy}
\rho_{\chi} (t) \simeq \frac{M^2\chi^2_*}{2}  \cdot \left(\frac{a_*}{a (t)} \right)^3 \; .
\end{equation}
Note that the estimate~\eqref{init} is applicable to both cases with the particles $\phi$ being relativistic or non-relativistic at the moment of the inverse phase transition. In the former case, $\eta^2 (T) $ is given by Eq.~\eqref{tension} and hence $\kappa=1$. 
In the latter case, $\kappa \simeq m_{\phi}/T_{sym} \simeq 1$. Moreover, the estimate~\eqref{init} works also in the situations, when the initial offset of the field $\chi$ has a non-thermal origin. Namely, other DM couplings, to the Ricci scalar or primordial magnetic fields 
discussed in Refs.~\cite{Babichev:2020xeg, Ramazanov:2020ajq}, respectively, lead to the same dependence of $\chi_*$ on the mass $M$ and the Hubble rate $H_*$ modulo the order one factor $\kappa$. 

\begin{figure}[tb!]
  \begin{center}
    \includegraphics[width=0.32\columnwidth,angle=0]{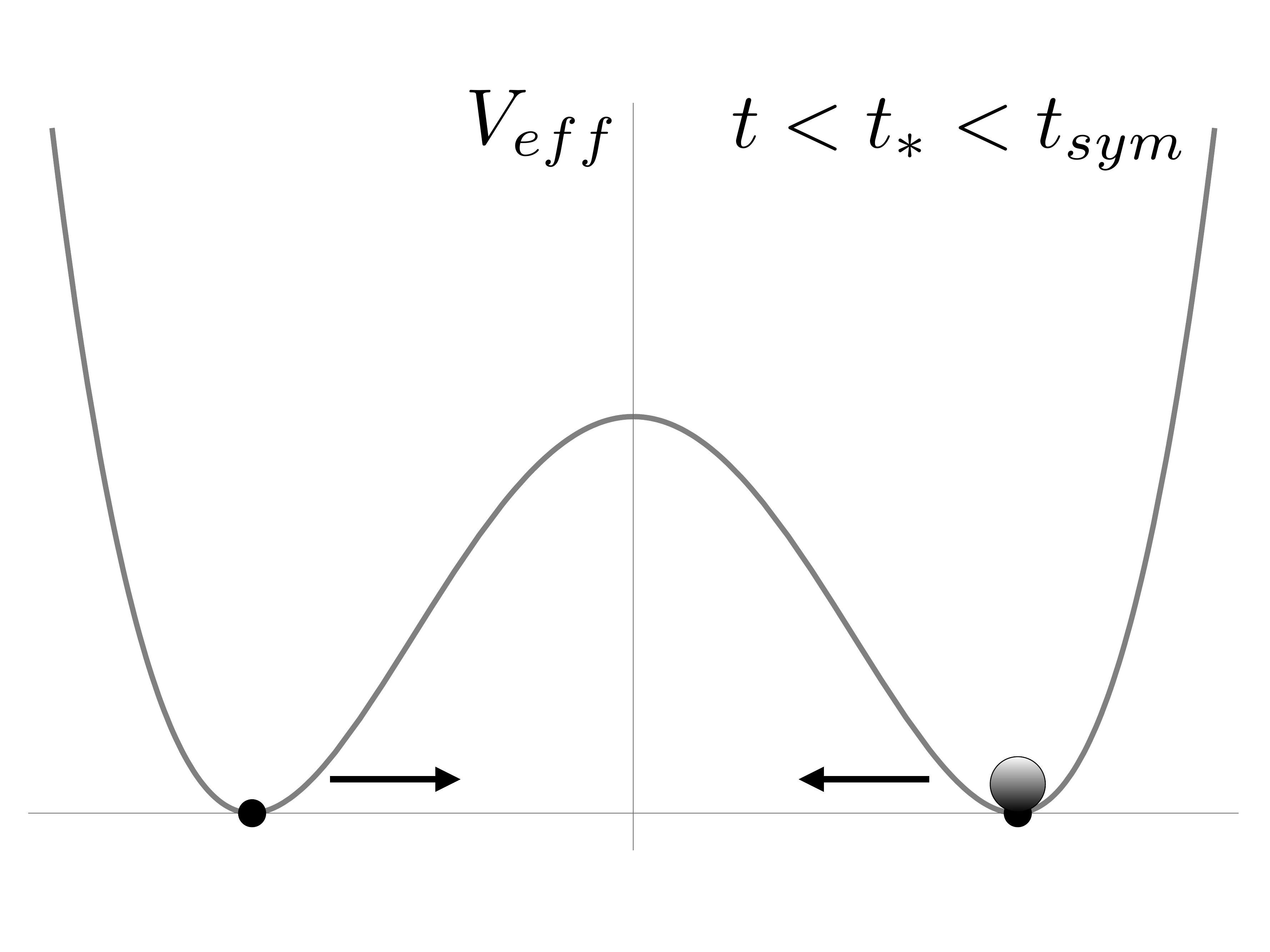}
        \includegraphics[width=0.32\columnwidth, angle=0]{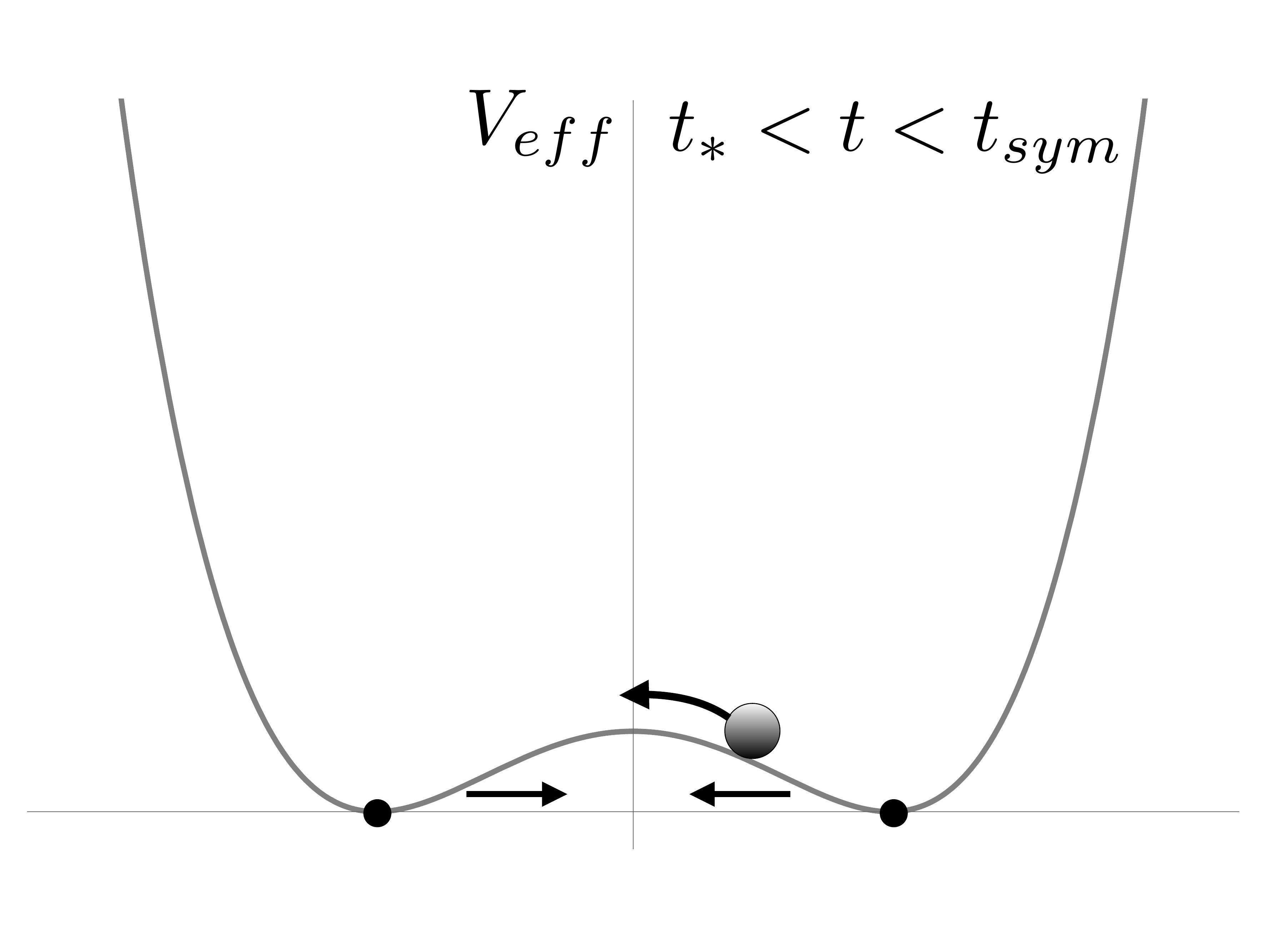}
          \includegraphics[width=0.32\columnwidth, angle=0]{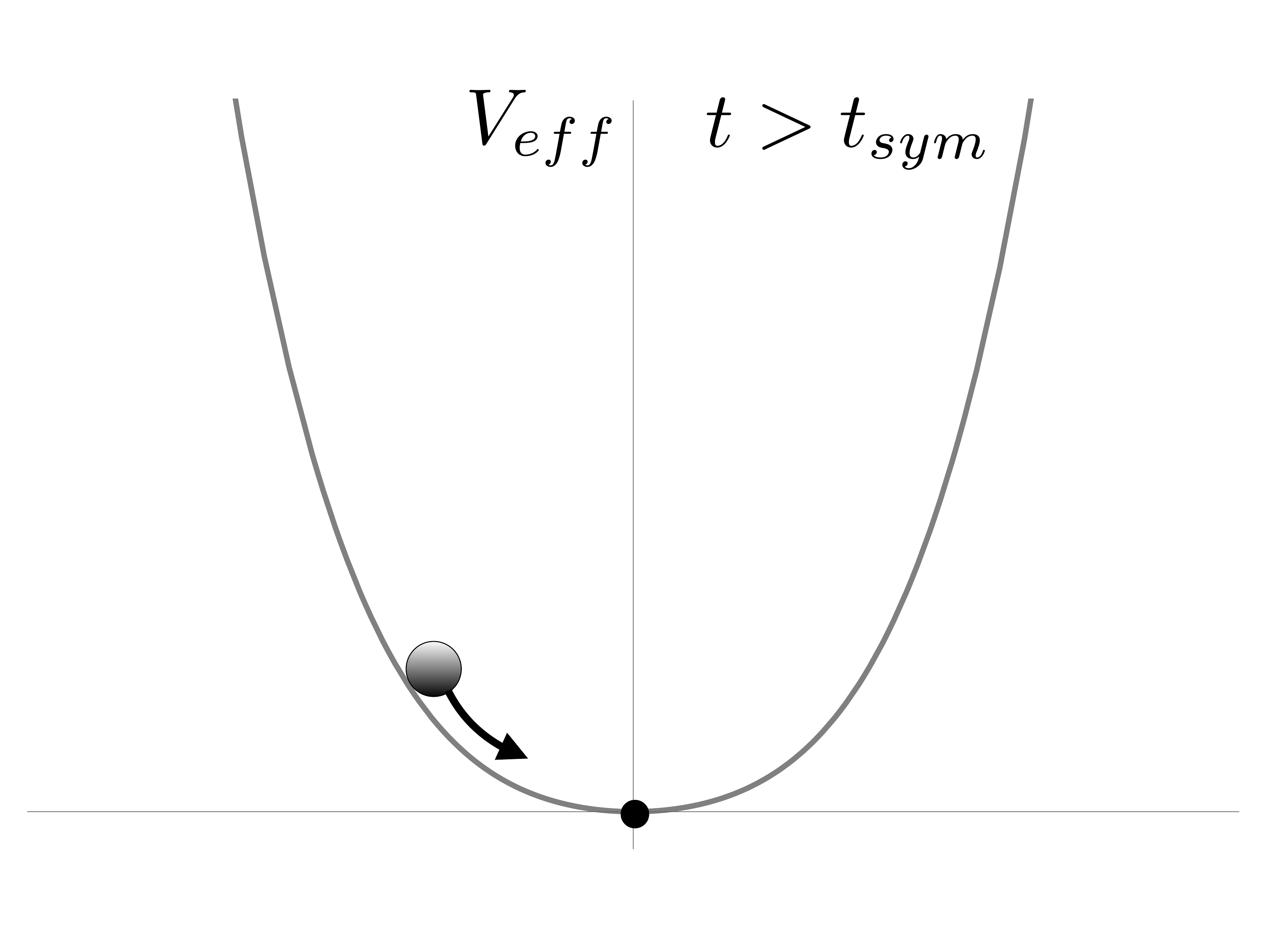}
  \caption{DM production at the inverse phase transition is shown. One assumes that the field $\chi$ is set to the minimum~\eqref{min} of its effective potential well before the symmetry is restored (left figure). 
  The temperature-dependent minimum~\eqref{min} 
  slowly drifts towards zero. At $ t_{sym}>t>t_*$, the field $\chi$ is offset from the minimum and starts oscillating, while the symmetry is still spontaneously broken (figure in the middle). 
  At later times, $t>t_{sym}$, the symmetry is restored, and the field $\chi$ continues oscillating (right figure).}\label{ipt}
  \end{center}
\end{figure}

\begin{figure}[tb!]
  \begin{center}
    \includegraphics[width=\columnwidth,angle=0]{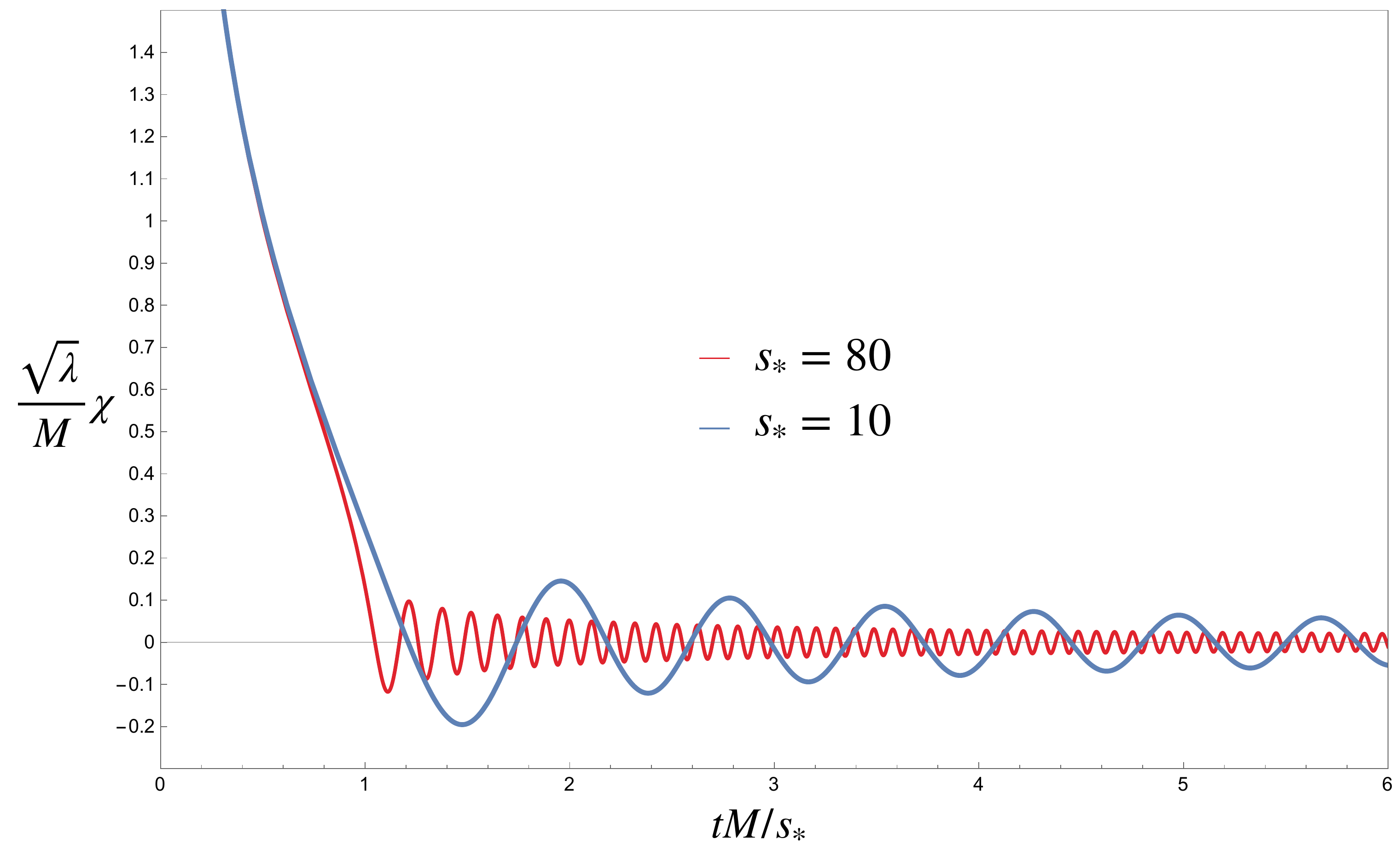}
  \caption{Numerically obtained evolution of the scalar field $\chi$ is plotted in rescaled variables $\bar{\chi}=\chi\sqrt{\lambda}/M$ and $s=t\,M/s_{*}$ for two different values of the dimensionless parameter $s_{*} \simeq M/(2H_*)$ (see the exact definition~\eqref{def}). In these rescaled variables, the moment of symmetry restoration $t_{sym}$, when the thermal mass of the field $\chi$ becomes equal to its bare mass, always corresponds to $s=1$. The larger $s_{*}$ is, the more accurately does $\chi$ track the minimum of the effective potential \eqref{spontaneous}, before the symmetry gets restored at $t_{sym}$.}
  \label{Evolve}
  \end{center}
\end{figure}

We checked the estimate~\eqref{init} with numerical simulations assuming the scenario, where the particles $\phi$ are still relativistic at the moment of the inverse phase transition. In that case, the equation of motion for the field $\chi$ is given by 
\begin{equation}
\label{eqinverse}
\ddot{\chi}+3H\dot{\chi}+\left(M^{2}-\frac{g^{2}NT^{2}}{12}\right)\chi+\lambda\chi^{3}=0 \; .
\end{equation}
The numerical solution of this equation with the initial condition at the times $t \ll t_{sym}$ given by Eq.~\eqref{onset} is shown in Fig.~\ref{Evolve}. Note that Eq.~\eqref{eqinverse} can be written in the minimalistic form 
\begin{equation}
\frac{1}{s^2_{*}}\left(\ddot{\bar{\chi}}+\frac{3}{2s}\dot{\bar{\chi}}\right)+\left(1-\frac{1}{s}\right)\bar{\chi}+\bar{\chi}^{3}=0 \; ,
\end{equation}
in terms of dimensionless field $\bar{\chi}=\chi\sqrt{\lambda}/M$, dimensionless time $s=t\,M/s_{*}$ and with the only single free parameter $s_{*}$ defined by 
\begin{equation}
\label{def}
s_{*}=\frac{g^{2}N}{24\pi}\sqrt{\frac{90}{g_{*} (T_*)}}\frac{M_{Pl}}{M} \simeq \frac{M}{2H_{*}} \; .
\end{equation}
 In the rescaled variables, the moment of time $t_{sym}$ when the symmetry is restored, always corresponds to $s=1$, while the amplitude at the onset of oscillations reads
 \begin{equation}
 \label{numerics}
 \bar{\chi}_{*} \simeq \frac{1}{6^{1/2} \cdot s^{1/3}_{*}} \; .
 \end{equation} 
 This is in an excellent agreement with the analytical estimate~\eqref{init}.

If the field $\chi$ constitutes all of DM in the Universe, the following condition must be fulfilled at the matter-radiation equality:  
\begin{equation}
\rho_{\chi} (t_{eq})  \simeq \rho_{rad} (t_{eq}) \simeq \frac{\pi^2 g_* (T_{eq}) T^4_{eq}}{30} \; ,
\end{equation} 
where $\rho_{rad} (t_{eq})$ is the radiation energy density at the equality time $t_{eq}$. From the latter, assuming that the inverse phase transition occurs at the radiation-dominated stage, using Eqs.~\eqref{hubble},~\eqref{energy}, ~\eqref{numerics}, 
and $T_{*} \approx T_{sym}$, we get the relation between the mass $M$ and the temperature $T_{sym}$:
\begin{equation}
\label{masstemp}
M \simeq 30~\mbox{eV} \cdot \frac{\beta^{3/10}}{\kappa^{1/5}} \cdot \left(\frac{g}{10^{-8}} \right)^{6/5}  \cdot \left(\frac{T_{sym}}{10~\mbox{GeV}} \right)^{1/2} \cdot \left(\frac{g_* (T_{sym})}{100} \right)^{1/5} \; .
\end{equation}
This relation is applicable to both scenarios with heavy and light particles $\phi$ defined by Eqs.~\eqref{h} and~\eqref{l}, respectively. In the former case, Eq.~\eqref{masstemp}, where one should just replace $T_{sym}$ by $m_{\phi}$ according to Eq.~\eqref{heavy}, yields 
the abundance constraint on the model parameter space. In the latter case, we obtain the analogous constraint equation on the model parameters substituting Eq.~\eqref{light} into Eq.~\eqref{masstemp}:
\begin{equation}
\label{constr2}
M \simeq 25~\mbox{eV} \cdot \frac{\beta^{3/5}}{\sqrt{N}} \cdot \left(\frac{g_* (T_{sym})}{100} \right)^{2/5} \cdot \left(\frac{g}{10^{-8}} \right)^{7/5} \; .
\end{equation}
The expression for the temperature at symmetry restoration $T_{sym}$ in that case is given by 
\begin{equation}
\label{symlight}
T_{sym} \simeq 9~\mbox{GeV} \cdot \frac{\beta^{3/5}}{N} \cdot \left(\frac{g_* (T_{sym})}{100} \right)^{2/5} \cdot \left(\frac{g}{10^{-8}} \right)^{2/5} \; .
\end{equation}
Taking the coupling $g \gtrsim 10^{-18}$ to have low frequency GWs still testable with PTAs, setting $\beta \simeq 1$, $N \simeq
10$, and $g_* (T_{sym}) \simeq 3.4$, we find that Eq.~\eqref{symlight} constrains the temperature at symmetry restoration $T_{sym} \gtrsim 20~\mbox{keV}$. Given also that the
generated DM out of the scalar field $\chi$ is supercold like an axion, we do not expect any departures from the canonical cold DM
scenario even for very small $\beta/N^2$ (except for the scales
$\sim\sqrt{HM}$ typical for generic scalar DM), in contrast to the scenario of the previous Subsection.

One important remark is in order here. In the scenario with light
particles $\phi$ there are also constraints from BBN. The temperature
$T_{sym}$ below $1~\mbox{MeV}$ corresponds to $g \lesssim 4 \cdot 10^{-17} N^{5/2}/\beta^{3/2}$. Having relativistic particles at these low temperatures is in conflict with BBN, if they are thermalized with the primordial plasma. The conflict can be avoided, if the particles 
$\phi$ departed from thermal equilibrium well before the inverse phase
transition. No such problem occurs in the scenario with heavy
particles $\phi$, where $T_{sym} \simeq m_{\phi}$, and it is typically 
well above the BBN temperature range.

We see that DM production at the inverse phase transition does not imply a lower bound on the ratio $\beta/N^2$; hence GWs emitted at the domain wall formation can be much more energetic than in the scenario of Subsection 6.1. 
The set of predictions regarding GWs is given by Eqs.\,\eqref{today} and~\eqref{frequency}. If the signal of interest is detected in the future, one will be able to infer values of constants $g$ and $\lambda$. Furthermore, knowing DM couplings, one can fix  the 
DM mass $M$, at least in the scenario with relativistic particles $\phi$ at the inverse phase transition.

Finally, let us note that DM production via the inverse phase transition is also possible in the Higgs portal scenario. In this case, however, one should fulfill the constraint $M \gtrsim gv_{SM}$, where $v_{SM} \approx 246~\mbox{GeV}$ is the Higgs 
field expectation value. See the discussion in the end of the previous Subsection. This imposes a strong limit on the temperature $T_{sym}$: 
\begin{equation}
\label{Higgstemp}
T_{sym} \gtrsim 60~\mbox{TeV} \cdot  \frac{\kappa^{2/5}}{\beta^{3/5}} \cdot \left(\frac{10^{-8}}{g} \right)^{2/5} \cdot \left(\frac{100}{g_* (T_{sym})} \right)^{2/5} \; .
\end{equation}
In the range of model parameters interesting for GW experiments, such large temperatures $T_{sym}$ cannot be reached in either scenario with particles $\phi$ (Higgs bosons) being non-relativistic or relativistic at the inverse 
phase transition. In the former case, $T_{sym} \simeq 100~\mbox{GeV}$, which is clearly inconsistent with Eq.~\eqref{Higgstemp}, unless one chooses very large values of $g \gtrsim 10^{-3}$ and/or $\beta \gtrsim 10^5$. In the latter case, $T_{sym}$ is even smaller, as it follows from Eq.~\eqref{symlight}. Consequently, we do not expect observable GWs in the DM production scenario involving the Higgs portal.

\subsection{Freeze-in mechanism}

Perhaps, the most well-known way of generating feebly coupled DM is given by the freeze-in mechanism~\cite{McDonald:2001vt, Hall:2009bx}. As in Section\,\ref{subsec:inversed-2order}, here we assume that large oscillations of the field $\chi$ produced at the phase transition fully decay in one or another way. The decay can be organized as it is
described in Appendix~B. In this case, the observed relic abundance
of DM can be reached via the freeze-in: by annihilations of (beyond)
the SM particles (the particles $\phi$ in our case). The DM number
density $n_\chi$ grows with time according to the equation:
\begin{equation}
\frac{dn_{\chi}}{dt}+3H n_{\chi} \approx \sum_i \langle \sigma_{\phi_i \phi_i \rightarrow \chi \chi} v \rangle \cdot n^{2}_{\phi_i, eq} \; ,
\end{equation}
where the sum is over the degrees of freedom constituting the field $\phi$. As the particles $\phi$ are in thermal equilibrium, we have $n_{\phi_i, eq} \propto T^3$. Then, using $H(T) \propto T^2 \propto t^{-1} $ and $\langle \sigma_{\phi_i \phi_i \rightarrow \chi \chi} v\rangle \propto T^{-2}$ (the temperatures $T \gtrsim M$ are assumed), we get the temperature dependence of the number density of $\chi$-particles, $n_{\chi} \propto T^2$. Thus, the DM yield $Y_{DM}$ (the ratio of DM number density to the entropy density) behaves with the 
temperature as $Y_{DM} \propto 1/T$ and is independent of the mass $M$
in the ultrarelativistic regime, 
for $T \gg M$. The DM production terminates by kinematics, when the temperature $T$ drops below the mass $M$; at this point the DM yield picks the value $Y_{DM} \propto 1/M$, and 
since then remains constant. We conclude that the observed abundance of DM, $\Omega_{DM} 
\propto M \cdot Y_{DM}$, is independent of the mass $M$. 

As it follows, the relic abundance of DM is determined only by the constant $g$. A thorough analysis of DM production in the case of the Higgs portal coupling gives the value $g \approx 
3 \cdot 10^{-6}$~\cite{Chu:2011be, Yaguna:2011qn, Lebedev:2019ton, Bringmann:2021sth}. Generalization to the case of arbitrary $N$ gives\footnote{This discussion assumes $M \gtrsim m_{\phi}$. In the opposite case $M \lesssim m_{\phi}$, the freeze-in occurs at the temperature $T \sim m_{\phi}$. Hence, for the same $g$, DM is underproduced by the factor $\sim M/m_{\phi}$. 
To maintain the observed amount of DM, one should increase the
constant $g$ by the factor $(m_{\phi}/M)^{1/4}$ (recall that the cross
section is proportional to $g^4$). 
In the particular example of the Higgs portal coupling, one has $g \simeq 2 \cdot 10^{-6} \left(\frac{\mbox{GeV}}{M} \right)^{1/4}$ for $M \lesssim m_{\phi}$~\cite{Lebedev:2019ton}.}:
\begin{equation}
\label{freezein}
g \approx \frac{5 \cdot 10^{-6}}{N^{1/4}}  \; .
\end{equation}
The peak frequency \eqref{frequency} of GWs emitted by the domain wall network is thus fixed to be 
\begin{equation}
f_{ peak}(t_0) \simeq \frac{2 \cdot 10^5~\mbox{Hz} \cdot N^{1/4}}{B^{1/2}} \; . 
\end{equation}
GWs with such high frequencies are beyond the range of sensitivity of currently planned detectors. See, however, Refs.~\cite{Domcke:2020yzq, Aggarwal:2020olq} investigating prospects in this direction. 
Note that GWs emitted at the times $t \gg t_i$ have much lower frequencies compared to our estimate~\eqref{frequency}. For $T(t)/T_i \sim 10^{-3}-10^{-4}$ 
the frequency falls in the observable range. However, this comes at
the price of decreasing the peak energy density of GWs by $6-8$ orders of magnitude compared to our estimate~\eqref{today}.

The discussion above assumes that domain walls are formed at radiation-dominated stage. However, for the constant $g$ as in Eq.~\eqref{freezein}, the reheating temperature can be well below the temperature of domain wall 
formation naively estimated by Eq.~\eqref{initial}. In particular, this is the case in Starobinsky inflation~\cite{Starobinsky:1980te}, where the Universe reheats to $T_{reh} \simeq 3 \cdot 10^{9}~\mbox{GeV}$~\cite{Gorbunov:2010bn}. In this case, the set of predictions regarding GWs is given by Eqs.~\eqref{gwreheating} and~\eqref{frreheating}. If one assumes Starobinsky inflation, the peak frequency of GWs is in the range accessible by Einstein Telescope or Cosmic Explorer: 
\begin{equation}
f_{peak}(t_0) \simeq 70~\mbox{Hz} \; .
\end{equation}
At the same time, the energy density of GWs is estimated as 
\begin{equation}
\Omega_{gw, peak} h^2 \simeq \frac{2 \cdot 10^{-19}  \cdot N^{7/2}}{\beta^2} \cdot \left(\frac{100}{g_* (T_{reh})} \right)^{4/3}\; .
\end{equation}
This is too little to be observed, at least with currently planned facilities, unless $N$ is chosen to be very large, $N ={\cal O} (100)$. According to the discussion in Section~5, that conclusion may be altered in more exotic preheating scenarios involving the long stage described by a stiff equation of state.

\section{Scenario with surviving domain walls}

So far we have assumed that the masses of the fields $\chi$ and $\phi$ satisfy $M^2>0$ and $m^2_{\phi}>0$. With such a choice of signs, the inverse phase transition inevitably happens leading to the dissolution of the domain wall network. 
On the other hand, if $M^2<0$, $Z_2$-symmetry remains spontaneously broken and domain walls survive until present with the tension stabilized at
\begin{equation}
\sigma_{wall} \simeq \frac{2\sqrt{2}}{3} \cdot \frac{|M|^3}{\lambda} \; .
\end{equation}
The limit on $\sigma_{wall}$ deduced from an accurate analysis of domain walls effect on the cosmic microwave background~\cite{Lazanu:2015fua} reads
\begin{equation}
\label{bound}
\sigma_{wall} <\left( 0.93~\mbox{MeV} \right)^3\quad (95\%~\mbox{CL})  \; . 
\end{equation}
This bounds the mass $|M|$ to be 
\begin{equation}
\label{limitmass}
|M| \lesssim 2\cdot 10^{-5}~\mbox{eV} \cdot \beta^{1/3} \cdot \left(\frac{g}{10^{-8}} \right)^{4/3} \; .
\end{equation}
Notably, the upper bound in Eq.~\eqref{bound} is very close to the rough limit $\sigma_{wall} \lesssim (\mbox{MeV})^3$ following from the requirement that domain walls do not overclose the Universe.

Again one can consider the field $\chi$ for the role of DM with relic abundance produced at the second order phase transition, as it is discussed in Subsection 6.1. Generically, DM stability is protected by symmetries. Thus, one may expect the DM stability to be compromised in the scenario 
with spontaneously broken $Z_2$-symmetry. Nevertheless, DM decay can be forbidden or strongly suppressed, if the DM field couples only to heavier particles. 
From Eq.~\eqref{limitmass} and the relation~\eqref{massdirect}, we get the bounds on the temperature $T_{dm}$ at the onset of DM phase: 
\begin{equation}
\label{dwtemperature}
70~\mbox{eV} \lesssim T_{dm} \lesssim \frac{5~\mbox{keV} \cdot \beta^{1/3}}{N^{1/2}} \cdot \left(\frac{g_* (T_{i})}{g_* (T_{dm})} \right)^{1/3} \cdot \left(\frac{g}{10^{-8}} \right)^{1/3} \; .
\end{equation}
Recall that the lower bound here follows from the requirement that we do not overproduce dark radiation at BBN, see Eq.~(\ref{lowerbbn}). 
We see that for the model parameters interesting for GW experiments, i.e., $\beta \simeq 1$, $N \simeq 10$, and $g \lesssim 10^{-8}$, the temperature $T_{dm}$ is confined to a rather limited range. 

Finally, let us discuss, how masses of the particles $\phi$ interfere in this picture. If they are substantially heavy and $m^2_{\phi}>0$, the portal interaction effectively switches off, as soon as the particles $\phi$ become non-relativistic. This leads to symmetry restoration and domain wall dissolution, provided that dynamics of the field $\chi$ is dominated by the quartic potential at these times. Later on, the bare mass term of the field $\chi$ (with $M^2<0$) comes 
into play, the symmetry gets spontaneously broken again, and domain walls reappear. For $m^2_{\phi}<0$, the field $\phi$ has a non-zero expectation 
value $v_{\phi}=\frac{|m_{\phi}|}{\sqrt{\lambda}_{\phi}}$. In this case, the effective mass squared of the field $\chi$ at $\chi=0$ is given by $M^2_{eff}=M^2-g^2v^2_{\phi}$. Note that $M^2<0$ implies $|M_{eff}| \gtrsim gv_{\phi}$. Using Eq.~\eqref{limitmass}, where one should replace $M$ by $M_{eff}$, we get a bound on the expectation value $v_{\phi}$:
\begin{equation}
v_{\phi} \lesssim 2~\mbox{keV} \cdot \beta^{1/3} \cdot \left(\frac{g}{10^{-8}} \right)^{1/3} \; ,
\end{equation}
which is well below the expectation value of the SM Higgs field in the interesting range of parameter space. Furthermore, 
masses of the particles $\phi$ turn out to be very small in this scenario and contribute to dark radiation at the times of BBN. 
This option is excluded, unless one assumes that the particles $\phi$ are not in the equilibrium with primordial plasma at very early times. Thus, the scenario with $M^2<0$ and $m^2_{\phi}<0$ turns out to be quite restrictive, 
in the range of constants relevant for GW observations.

\section{Discussions}

The idea that a symmetry, which is spontaneously broken initially, gets restored
later, allows one to enjoy rich phenomenology emerging from domain walls and monopoles and at the same 
time avoid problems with overclosing the Universe. Long time ago an inverse phase transition has been employed to get rid of monopoles in the early Universe~\cite{Langacker:1980kd}. 
In this work we considered the symmetry restoration as the remedy for the domain wall problem~\cite{Vilenkin:1981zs}. There is a simple realization of this setup, which involves two scalar fields, one 
of them being in thermal equilibrium with plasma. The attractive feature of the two-field model is that it enjoys an approximate scale-invariance at high temperatures. As the temperature decreases, bare mass terms 
of the fields come into play, and the symmetry gets restored. Domain walls appearing in this scenario have the energy density decreasing faster than the energy density of the surrounding radiation. Therefore, they have no effect on the cosmological expansion of the Universe. Nevertheless, they can be "visible" through the emission of 
GWs. 

An accurate determination of GWs spectrum from melting domain walls is still
to be accomplished. We argue that such a study would be indeed interesting and important for two reasons.
First, the peak energy density of
GWs can be within the range of sensitivity of currently planned
detectors like Einstein Telescope, LISA, and SKA. Second, according to the estimates in Section~4, the spectrum of GWs emitted in our case, is distinguishable from the spectrum of GWs produced by constant tension domain walls, at least in the low-frequency part. This makes it possible to discriminate GW signatures of melting and constant tension domain walls in future detectors.

The range of model parameters interesting from the viewpoint of GW experiments corresponds to a very weakly coupled scalar field $\chi$ constituting domain walls. In this regard, the field $\chi$ shares similarities with an axion, at least in the low mass range, which is phenomenologically most interesting, cf. Eqs.~\eqref{massdirect} and~\eqref{constr2}. However, DM production mechanisms are different in two setups, despite that they both rely on a version of a misalignment mechanism. Compared to axions, which are offset from zero because they have small masses relative to the Hubble rate initially, the offset of the field $\chi$ is due to a large tachyonic mass originating from the thermal fluctuations of the second scalar\footnote{See Refs.~\cite{Batell:2021ofv, Chun:2021uwr} on thermal misalignment induced by the interaction with fermions.}. 
Note that formation of domain walls---albeit with a constant tension---is also common in a system with axions~\cite{Sikivie:1982qv}. In that case, however, GWs emitted are very weak in the range 
of parameter space, where DM is not overproduced~\cite{Hiramatsu:2012sc} (cf., Ref.~\cite{Gelmini:2021yzu})\footnote{The recent Ref.~\cite{Krajewski:2021jje} further challenges the possibility of getting observable GWs from metastable domain walls, which collapse due to a small explicit breaking of discrete symmetry.}. There is no such a limitation in our setup: two major ways of generating DM, based on the second order phase transition and on the inverse phase transition, are compatible with potentially observable GWs.

\section*{Acknowledgments}

The work of E.B. was supported by Theoretical Physics and Mathematics Advancement Foundation "BASIS" and the CNRS/RFBR Cooperation program
for 2018-2020 n.~1985 ``Modified gravity and black holes: consistent
models and experimental signatures''. D.~G. was supported by Russian Foundation for Basic Research grant 18-52-15001- NCNIa. S.~R. acknowledges the Czech
Science Foundation GA\v CR, project 20-16531Y. A. V. is supported by
the European Regional Development Fund (ESIF/ERDF) and the Czech
Ministry of Education, Youth and Sports (M\v SMT) through the Project
CoGraDS- CZ.02.1.01/0.0/0.0/15 003/0000437.

\section*{Appendix~A}

The goal of this Appendix is to express the energy density of GWs~\eqref{intermone} in terms of model constants and parameters describing evolution of the Universe at preheating. 
We start with rewriting it as follows: 
\begin{equation}
\label{interm}
\Omega_{gw, peak} \cdot h^2 \simeq  \frac{4 \cdot 10^{-12} \cdot T^2_{h} \cdot N^3}{g^2 \cdot B^{3-\frac{2}{\gamma_2}}\cdot \beta^2 \cdot M^2_{Pl}} \cdot \left(\frac{a_{reh}}{a_h} \right)^{4\gamma_2-4} \cdot \left(\frac{100}{g_* (T_{reh})} \right)^{4/3}\; .
\end{equation}
The temperature $T_h$ at the onset of the roll can be expressed through $T_{reh}$ and $H_h$ using Eqs.~\eqref{slopes} and~\eqref{eos}: 
\begin{equation}
T_h =T_{reh} \cdot \left(\frac{a_{reh}}{a_h} \right)^{\gamma_2}= T_{reh} \cdot \left(\frac{H_h}{H_{reh}} \right)^{\frac{2\gamma_2}{3(1+w)}} \; .
\end{equation}
Then, we make use of $H_h \simeq N^{1/2} g T_h/\sqrt{12}$, which defines the onset of the roll, and obtain 
\begin{equation}
\label{Th}
T_{h} \simeq T_{reh} \cdot \left[\frac{15 N g^2 M^2_{Pl}}{2 \pi^2 g_* (T_{reh}) T^2_{reh}} \right]^{\frac{\gamma_2}{3(1+w)  -2\gamma_2}} \; .
\end{equation}
Expressing the ratio of scale factors $a_{reh}/a_{h}$ in Eq.~\eqref{interm} through the ratio of temperatures, i.e., $a_{reh}/a_{h} \propto (T_{h}/T_{reh})^{1/\gamma_2}$, and substituting Eq.~\eqref{Th}, we can rewrite Eq.~\eqref{interm} in the form:  
\begin{equation}
\label{complicated}
\Omega_{gw, peak} h^2 \simeq \frac{4 \cdot 10^{-12} \cdot N^4}{B^{3-\frac{2}{\gamma_2}} \cdot \beta^2} \cdot \left(\frac{3}{40 \pi^2} \right)^{\frac{6\gamma_2-4}{d(\gamma_2, w)}} \cdot \left(\frac{T_{reh}}{g N^{1/2} M_{Pl}} \right)^{\frac{s(\gamma_2, w)}{d(\gamma_2, w)}} 
\cdot   \left(\frac{100}{g_* (T_{reh})} \right)^{\frac{r(\gamma_2, w)}{3d(\gamma_2, w)}} \; ,
\end{equation}
where 
\begin{equation}
s(\gamma_2, w)=6w-16\gamma_2+14 \; , \quad d(\gamma_2, w)=3(1+w)-2\gamma_2 \; , \quad r(\gamma_2, w)=12w +10\gamma_2 \; .
\end{equation}
We crosscheck this formula by setting $w=1/3$ and $\gamma_2=1$. We obtain Eq.~\eqref{today}, if we assume the same values of $\tilde{\epsilon}_{gw}$ and ${\cal A}$.

Note that the coefficient $B$, which takes into account finite duration of the roll, is different compared to Eq.~\eqref{A} assuming that the roll occurs during the radiation-dominated stage. 
The coefficient $B$ can be obtained following the same steps as in Section~3. The field $\chi$ approaches the minimum according to (cf. Eq.~\eqref{greatsolution}) 
\begin{equation}
\tilde{\chi} (\tau) \simeq \frac{1}{2} \tilde{\chi}_h \cdot \mbox{exp} \left[\int^{\tau}_{\tau_h} |\tilde{M}_{eff} (\tau') | d\tau' \right] \; .
\end{equation}
Recall that $\tilde{\chi} \equiv \chi a$. The scale factor behaves with conformal time as $a(\tau) \propto \tau^{\frac{2}{1+3w}}$. Using that $|\tilde{M}_{eff} (\tau)| \propto T a \propto 1/a^{\gamma-1}$ and $|M_{eff, h}| \simeq H_h$, one gets for the integral in the exponent: 
\begin{equation}
\int^{\tau}_{\tau_h} |\tilde{M}_{eff} (\tau') | d\tau' \simeq \frac{2}{3+3w-2\gamma_2} \cdot \left(\frac{\tau_i}{\tau_h} \right)^{\frac{3+3w-2\gamma_2}{1+3w}} \; .
\end{equation}
From the condition $\chi_i \simeq \eta_i$, we obtain the equation defining $\tau_i/\tau_h$:
\begin{equation}
\frac{\chi_h}{2\eta_h} \cdot \left(\frac{\tau_h}{\tau_i} \right)^{\frac{2(1-\gamma_2)}{1+3w}} \cdot \mbox{exp} \left[\frac{2}{3+3w-2\gamma_2} \cdot \left(\frac{\tau_i}{\tau_h} \right)^{\frac{3+3w-2\gamma_2}{1+3w}} \right] \simeq 1 \; .
\end{equation}
With a logarithmic accuracy, the coefficient $B$ is given by
\begin{equation}
\label{complicatedA}
B \equiv \left(\frac{T_h}{T_i} \right)^2 \simeq \left(\frac{3+3w-2\gamma_2}{2} \right)^{\frac{4\gamma_2}{{3+3w-2\gamma_2}}} \cdot \left( \ln \frac{2\eta_h}{\chi_h}\right)^{\frac{4\gamma_2}{3+3w-2\gamma_2}} \; . 
\end{equation}
Again we crosscheck this formula by setting $w =1/3$ and $\gamma_2 =1$. We obtain Eq.~\eqref{A}, as it should be. 

When deriving Eqs.~\eqref{complicated} and~\eqref{complicatedA}, we did not take into account the coupling of the field $\chi$ to the Ricci scalar, which is plausible 
for $\xi \simeq 1$. Otherwise, this coupling can be relevant and the estimates~\eqref{complicated} and~\eqref{complicatedA} should be revisited. Note here an interesting possibility that the non-minimal coupling to gravity itself can drive spontaneous symmetry breaking, even 
when the temperature effects are negligible~\cite{Bettoni:2019dcw}. This happens, if the Ricci scalar takes a positive value for a sufficiently long period of time, e.g., during the kination stage.

\section*{Appendix~B}

Mechanisms of DM production via the inverse phase transition and by the freeze-in require that the large oscillations~\eqref{osc} following the direct phase transition decay efficiently, so that the field $\chi$ quickly relaxes to the minimum of the effective 
potential~\eqref{spontaneous}. The goal of this Appendix is to give an example of such a decay. For this purpose, let us extend the model~\eqref{basic} by means of another real scalar field $\varphi$, which is also not in thermal equilibrium 
with the hot plasma, described by the Lagrangian
\begin{equation}
{\cal L}_{\varphi}=\frac{(\partial_{\mu} \varphi)^2}{2}-\frac{\tilde{g}^4 \varphi^2 \chi^2}{2} +\frac{\lambda_{\varphi} \varphi^4}{4}\; .
\end{equation}
We assume that $|\tilde{g}^4| \lesssim g^4$ and neglect the quartic interaction of the field $\varphi$ in what follows as well as its possible bare mass. Evolution of the modes ${\bf k}$ of the redefined field $\tilde{\varphi}=\varphi \cdot a$ in the presence of the oscillating field $\chi$ is described by the equation:
\begin{equation}
\label{massless}
\tilde{\varphi}''_{{\bf k}}+ \left[k^2+\tilde{g}^4 \eta^2 a^2 +2\tilde{g}^4 \eta \delta \chi a^2 \sin \left(\tilde{M}_{eff} \tau \right)  \right] \tilde{\varphi}_{{\bf k}} =0 \; .
\end{equation}
Prime denotes the derivative with respect to the conformal time $d\tau =dt/a(t)$; recall that $\tilde{M}_{eff}=M_{eff} a=\sqrt{2\lambda} \eta a$. To make the analysis tractable, 
we neglected the quadratic term in $\delta \chi$ in Eq.~\eqref{massless}. However, keeping the quadratic term instead of linear one does not alter the picture qualitatively. 
Therefore, it is unlikely that mixing both terms dramatically changes
the analysis. Note that Eq.~\eqref{massless} has the form of Mathieu
equation:
\begin{equation}
\label{mathieu}
\frac{d^2y}{dz^2}+\left[C+2q \sin (2z) \right]y=0 \; .
\end{equation}
This becomes clear upon the identification: 
\begin{equation}
y=\tilde{\varphi}_{{\bf k}}=\varphi_{{\bf k}} a \qquad z=\frac{\tilde{M}_{eff} \tau}{2} \qquad C=\frac{4k^2}{\tilde{M}^2_{eff}}+\frac{2}{\tilde{\beta}} \qquad q =\frac{2 \delta \chi}{\tilde{\beta} \eta} \; ,
\end{equation}
where $\tilde{\beta} \equiv \lambda/\tilde{g}^4$; note that $|\tilde{\beta}| \gtrsim \beta \gtrsim 1$. We used that $M_{eff}=\sqrt{2\lambda} \eta$. Mathieu equation leads to the instability manifested as the exponential growth of the function $y$. Namely, 
for $C \approx 1$ and $q\ll 1$, one has $y \propto \mbox{exp} \left(\frac{|q|z}{2} \right)$. Getting back to the original variables, 
we conclude that 
\begin{equation}
\tilde{\varphi}_{{\bf k}} \propto \mbox{exp} \left(\frac{\delta \chi \tilde{M}_{eff} \tau}{2\tilde{\beta} \eta} \right) \; ,
\end{equation} 
for $k \approx \tilde{M}_{eff}/2$. Since we are 
in the regime $M_{eff} \gg H$, transfer of energy from the field $\chi$ oscillations to the field $\varphi$ occurs within a few Hubble times, at least for not extremely large $\tilde{\beta}$. This is analogous to what one has at preheating stage in the narrow parametric resonance regime~\cite{Kofman:1997yn}. In our case, the resonant growth of the field $\varphi$ modes is particularly efficient due to an approximate scale-invariance of our scenario (at very high temperatures)~\cite{Greene:1997fu}.

One may worry that oscillations of the field $\chi$ following the inverse phase transition may also decay 
into the particles $\varphi$. However, this decay is much less efficient, in particular because of the lost scale-invariance. Evolution of the modes of the field $\varphi$ is again described by Eq.~\eqref{mathieu}, now with
\begin{equation}
y=\varphi_{{\bf k}} a^{3/2} \qquad  z=Mt \qquad C \approx \frac{k^2}{M^2 a^2} \qquad q \approx \frac{7 \cdot 10^{-2}}{\tilde{\beta}} \cdot \left(\frac{H_{*}}{M} \right)^{2/3} \cdot \left(\frac{a_*}{a(t)} \right)^3 \; .
\end{equation}
We see that the growth of the particular $k$-mode is reduced with time because of the Universe expansion. This growth is most efficient at $t \simeq t_*$ and its rate 
is estimated as $\Gamma \simeq 3\cdot 10^{-2} M^{1/3} H^{2/3}_*/\tilde{\beta}$. 
In the range of parameters interesting for GW observations, $g \simeq 10^{-18}-10^{-8}$, one can always ensure 
$\Gamma \ll H_*$ by choosing $|\tilde{\beta} | \gtrsim 10-10^{3}$.


\begin{thebibliography}{99}

\bibitem{Zeldovich:1974uw}
Y.~B.~Zeldovich, I.~Y.~Kobzarev and L.~B.~Okun,
Zh. Eksp. Teor. Fiz. \textbf{67} (1974), 3-11
SLAC-TRANS-0165.

\bibitem{Kibble:1976sj}
T.~W.~B.~Kibble,
J. Phys. A \textbf{9} (1976), 1387-1398.

\bibitem{Gelmini:1988sf}
G.~B.~Gelmini, M.~Gleiser and E.~W.~Kolb,
Phys. Rev. D \textbf{39} (1989), 1558

\bibitem{Coulson:1995nv}
D.~Coulson, Z.~Lalak and B.~A.~Ovrut,
Phys. Rev. D \textbf{53} (1996), 4237-4246

\bibitem{Lazarides:1982tw}
G.~Lazarides and Q.~Shafi,
Phys. Lett. B \textbf{115} (1982), 21-25.



\bibitem{Dvali:1995cc}
G.~R.~Dvali and G.~Senjanovic,
Phys. Rev. Lett. \textbf{74} (1995), 5178-5181
[arXiv:hep-ph/9501387 [hep-ph]].




\bibitem{Vilenkin:1981zs}
A.~Vilenkin,
Phys. Rev. D \textbf{23} (1981), 852-857.

\bibitem{Vilenkin:2000jqa}
A. Vilenkin, and E. P. S. Shellard, {\it Cosmic Strings and
Other Topological Defects}, Cambridge University Press (1994).

\bibitem{Weinberg:1974hy}
S.~Weinberg,
Phys. Rev. D \textbf{9} (1974), 3357-3378.





\bibitem{Ramazanov:2021eya}
S.~Ramazanov, E.~Babichev, D.~Gorbunov and A.~Vikman,
[arXiv:2104.13722 [hep-ph]].

\bibitem{Hiramatsu:2013qaa}
T.~Hiramatsu, M.~Kawasaki and K.~Saikawa,
JCAP \textbf{02} (2014), 031
[arXiv:1309.5001 [astro-ph.CO]].

\bibitem{Gleiser:1998na}
M.~Gleiser and R.~Roberts,
Phys. Rev. Lett. \textbf{81} (1998), 5497-5500
[arXiv:astro-ph/9807260 [astro-ph]].









\bibitem{Sakharov:2021dim}
A.~S.~Sakharov, Y.~N.~Eroshenko and S.~G.~Rubin,
Phys. Rev. D \textbf{104} (2021) no.4, 043005
[arXiv:2104.08750 [hep-ph]].

\bibitem{Press:1989yh}
W.~H.~Press, B.~S.~Ryden and D.~N.~Spergel,
Astrophys. J. \textbf{347} (1989), 590-604.

\bibitem{Garagounis:2002kt}
T.~Garagounis and M.~Hindmarsh,
Phys. Rev. D \textbf{68} (2003), 103506
[arXiv:hep-ph/0212359 [hep-ph]].



\bibitem{Leite:2012vn}
A.~M.~M.~Leite, C.~J.~A.~P.~Martins and E.~P.~S.~Shellard,
Phys. Lett. B \textbf{718} (2013), 740-744
[arXiv:1206.6043 [hep-ph]].

\bibitem{Baldes:2018emh}
I.~Baldes and C.~Garcia-Cely,
JHEP \textbf{05} (2019), 190
[arXiv:1809.01198 [hep-ph]].

\bibitem{Azatov:2021ifm}
A.~Azatov, M.~Vanvlasselaer and W.~Yin,
JHEP \textbf{03} (2021), 288
[arXiv:2101.05721 [hep-ph]].

\bibitem{Nakayama:2021avl}
K.~Nakayama and W.~Yin,
JHEP \textbf{10} (2021), 026
[arXiv:2105.14549 [hep-ph]].

\bibitem{Birrell}
N.~D.~Birrell and P.~G.~W.~Davies, {\it Quantum Fields in Curved Space}, Cambridge University Press (1982).

\bibitem{Chai:2020onq}
N.~Chai, S.~Chaudhuri, C.~Choi, Z.~Komargodski, E.~Rabinovici and M.~Smolkin,
Phys. Rev. Lett. \textbf{125} (2020) no.13, 131603

\bibitem{Chai:2020zgq}
N.~Chai, S.~Chaudhuri, C.~Choi, Z.~Komargodski, E.~Rabinovici and M.~Smolkin,
Phys. Rev. D \textbf{102} (2020) no.6, 065014
[arXiv:2005.03676 [hep-th]].

\bibitem{Chai:2021djc}
N.~Chai, A.~Dymarsky and M.~Smolkin,
Phys. Rev. Lett. \textbf{128} (2022) no.1, 011601
[arXiv:2106.09723 [hep-th]].

\bibitem{Dolan:1973qd}
L.~Dolan and R.~Jackiw,
Phys. Rev. D \textbf{9} (1974), 3320-3341.




\bibitem{Markkanen:2015xuw}
T.~Markkanen and S.~Nurmi,
JCAP \textbf{02} (2017), 008
[arXiv:1512.07288 [astro-ph.CO]].

\bibitem{Fairbairn:2018bsw}
M.~Fairbairn, K.~Kainulainen, T.~Markkanen and S.~Nurmi,
JCAP \textbf{04} (2019), 005
[arXiv:1808.08236 [astro-ph.CO]].

\bibitem{Bassett:1997az}
B.~A.~Bassett and S.~Liberati,
Phys. Rev. D \textbf{58} (1998), 021302
[erratum: Phys. Rev. D \textbf{60} (1999), 049902]
[arXiv:hep-ph/9709417 [hep-ph]].

\bibitem{Felder:2000hj}
G.~N.~Felder, J.~Garcia-Bellido, P.~B.~Greene, L.~Kofman, A.~D.~Linde and I.~Tkachev,
Phys. Rev. Lett. \textbf{87} (2001), 011601
[arXiv:hep-ph/0012142 [hep-ph]].






\bibitem{Emond:2021vts}
W.~T.~Emond, S.~Ramazanov and R.~Samanta,
[arXiv:2108.05377 [hep-ph]].












\bibitem{Hild:2010id}
S.~Hild \textit{et al.}
Class. Quant. Grav. \textbf{28} (2011), 094013
[arXiv:1012.0908 [gr-qc]].

\bibitem{Evans:2016mbw}
B.~P.~Abbott \textit{et al.} [LIGO Scientific],
Class. Quant. Grav. \textbf{34} (2017) no.4, 044001
[arXiv:1607.08697 [astro-ph.IM]].

\bibitem{Kawamura:2006up}
S.~Kawamura~\textit{et al.}
Class. Quant. Grav. \textbf{23} (2006), S125-S132.

\bibitem{TianQin:2015yph}
J.~Luo \textit{et al.} [TianQin],
Class. Quant. Grav. \textbf{33} (2016) no.3, 035010
[arXiv:1512.02076 [astro-ph.IM]].



\bibitem{Audley:2017drz}
P.~Amaro-Seoane \textit{et al.} [LISA],
[arXiv:1702.00786 [astro-ph.IM]].





\bibitem{Moore:2014eua}
C.~J.~Moore, S.~R.~Taylor and J.~R.~Gair,
Class. Quant. Grav. \textbf{32} (2015) no.5, 055004
[arXiv:1406.5199 [astro-ph.IM]].

\bibitem{gwplotter}
gwplotter.com

\bibitem{Moore:2014lga}
C.~J.~Moore, R.~H.~Cole and C.~P.~L.~Berry,
Class. Quant. Grav. \textbf{32} (2015) no.1, 015014
[arXiv:1408.0740 [gr-qc]].






\bibitem{Romano:2016dpx}
J.~D.~Romano and N.~J.~Cornish,
Living Rev. Rel. \textbf{20} (2017) no.1, 2
[arXiv:1608.06889 [gr-qc]].


\bibitem{Aghanim:2018eyx}
N.~Aghanim \textit{et al.} [Planck],
Astron. Astrophys. \textbf{641} (2020), A6
[arXiv:1807.06209 [astro-ph.CO]].



\bibitem{Garcia-Bellido:2021zgu}
J.~Garcia-Bellido, H.~Murayama and G.~White,
[arXiv:2104.04778 [hep-ph]].




\bibitem{Starobinsky:1980te}
A.~A.~Starobinsky,
Phys. Lett. B \textbf{91} (1980), 99-102.

\bibitem{Gorbunov:2010bn}
D.~S.~Gorbunov and A.~G.~Panin,
Phys. Lett. B \textbf{700} (2011), 157-162
[arXiv:1009.2448 [hep-ph]].

\bibitem{Giudice:2000ex}
G.~F.~Giudice, E.~W.~Kolb and A.~Riotto,
Phys. Rev. D \textbf{64} (2001), 023508
[arXiv:hep-ph/0005123 [hep-ph]].

\bibitem{McDonald:2001vt}
J.~McDonald,
Phys. Rev. Lett. \textbf{88} (2002), 091304
[arXiv:hep-ph/0106249 [hep-ph]].

\bibitem{Hall:2009bx}
L.~J.~Hall, K.~Jedamzik, J.~March-Russell and S.~M.~West,
JHEP \textbf{03} (2010), 080
[arXiv:0911.1120 [hep-ph]].





\bibitem{Chu:2011be}
X.~Chu, T.~Hambye and M.~H.~G.~Tytgat,
JCAP \textbf{05} (2012), 034
[arXiv:1112.0493 [hep-ph]].

\bibitem{Yaguna:2011qn}
C.~E.~Yaguna,
JHEP \textbf{08} (2011), 060
[arXiv:1105.1654 [hep-ph]].

\bibitem{Lebedev:2019ton}
O.~Lebedev and T.~Toma,
Phys. Lett. B \textbf{798} (2019), 134961
[arXiv:1908.05491 [hep-ph]].

\bibitem{Bringmann:2021sth}
T.~Bringmann, S.~Heeba, F.~Kahlhoefer and K.~Vangsnes,
[arXiv:2111.14871 [hep-ph]].

\bibitem{Yeche:2017upn}
C.~Y\`eche, N.~Palanque-Delabrouille, J.~Baur and H.~du Mas des Bourboux,
JCAP \textbf{06} (2017), 047
[arXiv:1702.03314 [astro-ph.CO]].


\bibitem{Kofman:1997yn}
L.~Kofman, A.~D.~Linde and A.~A.~Starobinsky,
Phys. Rev. D \textbf{56} (1997), 3258-3295
[arXiv:hep-ph/9704452 [hep-ph]].

\bibitem{Greene:1997fu}
P.~B.~Greene, L.~Kofman, A.~D.~Linde and A.~A.~Starobinsky,
Phys. Rev. D \textbf{56} (1997), 6175-6192
[arXiv:hep-ph/9705347 [hep-ph]].

\bibitem{Baker:2019ndr}
M.~J.~Baker, J.~Kopp and A.~J.~Long,
Phys. Rev. Lett. \textbf{125} (2020) no.15, 151102
[arXiv:1912.02830 [hep-ph]].

\bibitem{Dodelson:1989ii}
S.~Dodelson and L.~M.~Widrow,
Phys. Rev. Lett. \textbf{64} (1990), 340-343.

\bibitem{Dodelson:1989cq}
S.~Dodelson and L.~M.~Widrow,
Phys. Rev. D \textbf{42} (1990), 326-342.



\bibitem{Babichev:2020xeg}
E.~Babichev, D.~Gorbunov and S.~Ramazanov,
JCAP \textbf{08} (2020), 047
[arXiv:2004.03410 [hep-ph]].

\bibitem{Ramazanov:2020ajq}
S.~Ramazanov, F.~R.~Urban and A.~Vikman,
JCAP \textbf{02} (2021), 011
[arXiv:2010.03383 [astro-ph.CO]].





















\bibitem{Domcke:2020yzq}
V.~Domcke and C.~Garcia-Cely,
Phys. Rev. Lett. \textbf{126} (2021) no.2, 021104
[arXiv:2006.01161 [astro-ph.CO]].

\bibitem{Aggarwal:2020olq}
N.~Aggarwal~\textit{et al.}
[arXiv:2011.12414 [gr-qc]].

\bibitem{Lazanu:2015fua}
A.~Lazanu, C.~J.~A.~P.~Martins and E.~P.~S.~Shellard,
Phys. Lett. B \textbf{747} (2015), 426-432
[arXiv:1505.03673 [astro-ph.CO]].

\bibitem{Langacker:1980kd}
P.~Langacker and S.~Y.~Pi,
Phys. Rev. Lett. \textbf{45} (1980), 1.

\bibitem{Batell:2021ofv}
B.~Batell and A.~Ghalsasi,
[arXiv:2109.04476 [hep-ph]].

\bibitem{Chun:2021uwr}
E.~J.~Chun,
[arXiv:2109.07423 [hep-ph]].









\bibitem{Sikivie:1982qv}
P.~Sikivie,
Phys. Rev. Lett. \textbf{48} (1982), 1156-1159.

\bibitem{Hiramatsu:2012sc}
T.~Hiramatsu, M.~Kawasaki, K.~Saikawa and T.~Sekiguchi,
JCAP \textbf{01} (2013), 001
[arXiv:1207.3166 [hep-ph]].

\bibitem{Gelmini:2021yzu}
G.~B.~Gelmini, A.~Simpson and E.~Vitagliano,
Phys. Rev. D \textbf{104} (2021) no.6, 061301
[arXiv:2103.07625 [hep-ph]].

\bibitem{Krajewski:2021jje}
T.~Krajewski, J.~H.~Kwapisz, Z.~Lalak and M.~Lewicki,
[arXiv:2103.03225 [astro-ph.CO]].

\bibitem{Bettoni:2019dcw}
D.~Bettoni and J.~Rubio,
JCAP \textbf{01} (2020), 002
[arXiv:1911.03484 [astro-ph.CO]].











\end{thebibliography}
\end{document}